\documentclass{article}
\usepackage[utf8]{inputenc}
\usepackage[width=16cm,bottom=3cm]{geometry} 
\usepackage{dirtree}
\usepackage{enumitem}
\usepackage{graphicx}
\usepackage{subfiles}
\usepackage{tabularx}
\usepackage{natbib}
\usepackage{easy-todo}
\usepackage{chngcntr}
\usepackage{subcaption}
\usepackage{hyperref}

\counterwithin{figure}{section}
\counterwithin{table}{section}

\title{DOSE-I: A Multimodal Biosignal Dataset\\ of Procedural Sedation for Endoscopy}
\author{Jakob Garbe, Jan Kantelhardt, Thomas Schmid}
\date{}

\begin{document}

\begin{titlepage}

\newcommand{\HRule}{\rule{\linewidth}{0.5mm}} 

\centering
\includegraphics[width=13.5cm]{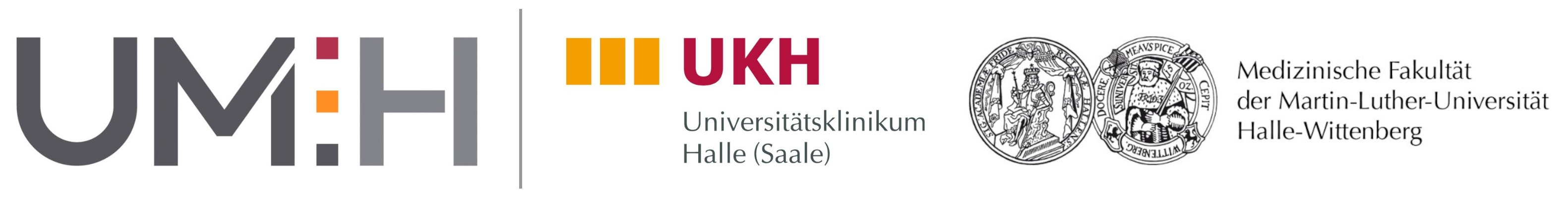}\\[0.25cm] 

\vspace{4.5cm}


\textsc{\LARGE Technical Report}\\[1.5cm] 

\makeatletter
\HRule \\[0.4cm]
{ \huge \bfseries \@title}\\[0.4cm] 
  {\large Dataset DOI: 10.5281/zenodo.18483292}\\[0.4cm]
\HRule \\[2.5cm]
 
{\large Jakob Garbe\\[0.15cm]Jan Kantelhardt\\[0.15cm]Katja Seeliger\\[0.15cm]Thomas Schmid}\\[2cm]


\makeatother


{\large Version 2}\\[0.75cm] 

{\large \today}\\[0.1cm] 

\vfill 

\end{titlepage}

\thispagestyle{empty} 
\vspace*{\fill}

\Large{
\noindent
\textbf{Universitätsmedizin Halle (Saale)}\\
}

\small
\noindent
\textbf{Universitätsklinikum Halle (Saale)}\\
Universitätsklinik und Poliklinik für Innere Medizin I\\[0.1cm]
\textbf{Martin‑Luther‑Universität Halle‑Wittenberg}\\
Medizinische Fakultät\\[0.1cm]
Ernst‑Grube‑Straße 40\\
D-06120 Halle (Saale)\\
Germany\\[0.5cm]
\begin{tabular}{@{} l l @{}}
    Contact: & Dr. Jakob Garbe\\
    Tel.: & +49 345 557 2928 \\
    E‑Mail: & \texttt{jakob.garbe1@uk-halle.de} \\
    Website: & \texttt{www.umh.de/innere1} \\[0.5cm]
    Project: & \texttt{https://safe-ai-research.github.io} \\
    Dataset:  & \texttt{https://safe-ai-research.github.io/DOSE-I/} 
\end{tabular}
\normalsize
\normalfont
\vspace*{0.1cm}

\newpage



\tableofcontents
\newpage

\addcontentsline{toc}{section}{List of Figures}
\listoffigures
\bigskip
\addcontentsline{toc}{section}{List of Tables}
\listoftables
\bigskip
\section*{Abbreviations}
\addcontentsline{toc}{section}{Abbreviations}
\begin{description}[align=right,labelwidth=3cm]
    \item[DOSE] Database of Sedation in Endoscopy
    \item[ECG] Electrocardiogram
    \item[EDG] Esophagogastroduodenoscopy
    \item[EEG] Electroencephalogram
    \item[LOC] Loss of Consciousness
    \item[MOAAS] Modified Observer’s Assessment of Alertness/Sedation Scale
    \item[NIBP] Non-invasive Blood Pressure
    \item[pEEG] Processed Electroencephalogram
    \item[PLETH] Photoplethysmogram
    \item[RESP] ECG-derived Respiration
    \item[ROC] Return of Consciousness
\end{description}

\newpage

\section*{Acknowledgements}
\addcontentsline{toc}{section}{Acknowledgements}
This work was supported by Deutsche Forschungsgemeinschaft (DFG, German Research Foundation) under Project number 547230187. 

The authors sincerely thank all colleagues and clinical staff who contributed to the acquisition and curation of the DOSE-I dataset. In particular, we would like to thank:
\begin{itemize}
    \item Florian D\"unninghaus (Warendorf, Germany) for his contribution in collection of the dataset;
    \item Josephin Damm (Halle, Germany), Karla Erffmeier (Marburg, Germany) and Leona M\"oller (Marburg, Germany) for annotation of EEGs and for their insights into signal quality;
    \item Andreas Schwab (Halle, Germany) for proof-reading of the dataset documentation.
\end{itemize}

\newpage

\section{Overview}

In this document, we describe characteristics and technical details of the multimodal biosignal dataset DOSE-I of procedural sedation for endoscopy \cite{garbe2026}. The DOSE-I dataset includes 78.5 hours of recording in 171 records ranging from 6.7 to 70.8 minutes (mean: 27.5, SD: 11.6) of 281 endoscopic procedures. 1129 (median: 6 per record) transitions of consciousness and 7328 (median: 39 per record) individual sedation depth were recorded using MOAA/S labels. 

In addition to clinically annotated biosignals, the DOSE-I dataset provides detailed static data about the respective study subject and metadata about the respective recordings. To further support future research, we provide details about artifact detection and preprocessed pEEG features, too. C code used for this preprocessing is provided separately via Github\footnote{https://safe-ai-research.github.io/DOSE-I}. The figure below gives an overview of the folder and file structure of the dataset available via Zenodo\footnote{DOI: 10.5281/zenodo.18483292}

\bigskip

\dirtree{%
.1 DOSE-I.
.2 Data\_Use\_Agreement.txt\DTcomment{conditions for using the dataset}.
.2 data\DTcomment{biosignal time series, numeric values, and clinical events}.
.3 10-XXX.csv\DTcomment{one file per record, total 171 records}.
.2 metadata\DTcomment{record length, labels, events, signals, and completeness}.
.3 DOSE-I\_metadata.csv\DTcomment{SoC and MOAAS counts, total propofol, tracks, errors}.
.3 DOSE-I\_metadata\_readme.md\DTcomment{description of metadata columns}.
.2 static\DTcomment{subject, medication, condition, and procedure data}.
.3 DOSE-I\_static\_data.csv\DTcomment{age, sex, ASA, BMI, medication, conditions, procedure}.
.3 DOSE-I\_static-data\_readme.md\DTcomment{description of static data columns}.
.2 artifacts\DTcomment{EEG artifact annotations}.
.3 DOSE-I\_artifacts\_readme.md\DTcomment{artifact detection and reason codes}.
.3 artifacts.
.4 10-XXX\_artifacts.csv\DTcomment{one artifact file per record, total 171 records}.
.4 artifact\_plots.
.5 DOSE-I\_artifact-plots\_X.png\DTcomment{artifact summary plots, total 6}.
.2 pEEG\DTcomment{EEG features with vital, drug, and label columns}.
.3 pEEG.
.4 10-XXX\_pEEG.csv\DTcomment{one feature file per record, total 171 records}.
.3 pEEG\_parameter\_description.txt\DTcomment{description of pEEG and added columns}.
.2 plots\DTcomment{record overview plots}.
.3 10-XXX\_plot.png\DTcomment{one plot per record, total 171 records}.
}

\bigskip
This technical documentation is intended to supplement a recent publication on applying data-driven artificial intelligence for predicting sedation depth in endoscopy \citep{garbe2026}. To this end, we would like to encourage the reader to cite this reference as the primary academic source for the DOSE-I dataset. Here, we describe the dataset's study cohort, biosignals characteristics, data quality, and data format in more detail.

The DOSE-I dataset has been published under a Creative Commons license. Using the dataset is subject to the data use agreement (cf. Appendix \ref{sec:data-use-agreement}).

\vfill
\pagebreak

\section{Study Cohort}




The DOSE-I dataset originates from a prospective single-center observational study conducted in Halle (Saale), Germany, between May 2019 and July 2021. Participants were enrolled and observed following the predefined study protocol registered under ID DRKS00016605 in the German Clinical Trials Register\footnote{\url{https://drks.de/search/en/trial/DRKS00016605}}. Entitled \emph{Improving Safety and Efficacy of Endoscopic Procedures – A Deep Machine Learning based Depth of Sedation Monitor}, the primary objective of our study was to provide an empirical foundation for investigations into whether modern machine learning approaches can be used to automatically estimate and continuously monitor the depth of sedation during endoscopic interventions.  Initial findings from this study have previously been reported by \citet{Garbe2020}.


During the two-year study, a total of 171 recordings from individual endoscopic interventions were collected during routine clinical care at the Endoscopy Unit of the University Hospital Halle. Each recording corresponds to one or more endoscopic procedures (e.g. gastroscopy and colonoscopy) and contains synchronized physiological and clinical information acquired during sedation and endoscopic examination. The recorded interventions covered several common diagnostic and interventional endoscopic procedures, including esophagogastroduodenoscopy, colonoscopies, endoscopic ultrasound examinations, bronchoscopies, and endobronchial ultrasound procedures. These procedures include purely diagnostic examinations as well as  interventions such as fine-needle biopsies and percutaneous endoscopic gastrostomies. Therefore, they differ substantially with respect to duration, invasiveness, patient discomfort, and sedation requirements, thereby increasing the diversity and clinical relevance of the dataset.

\begin{figure}[h!]
    \centering
    \begin{subfigure}[b]{0.49\textwidth}
        \centering
        \includegraphics[width=.91\linewidth,trim=0 15 0 45, clip=TRUE]{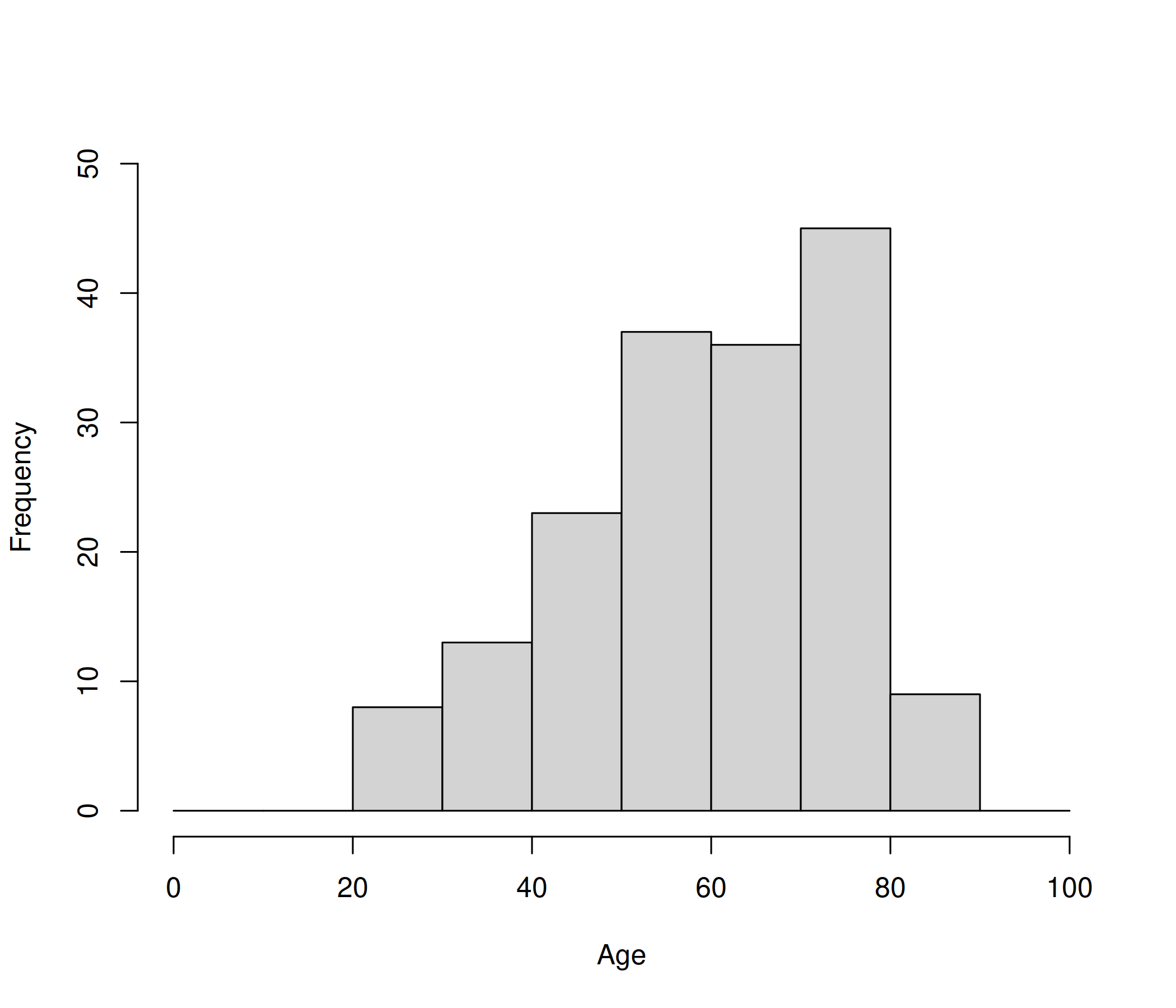}
        \caption{Age Distribution}
        \label{fig:sub2}
    \end{subfigure}
    \hfill
    \begin{subfigure}[b]{0.49\textwidth}
        \centering
        \includegraphics[width=.95\linewidth,trim=0 0 0 45, clip=TRUE]{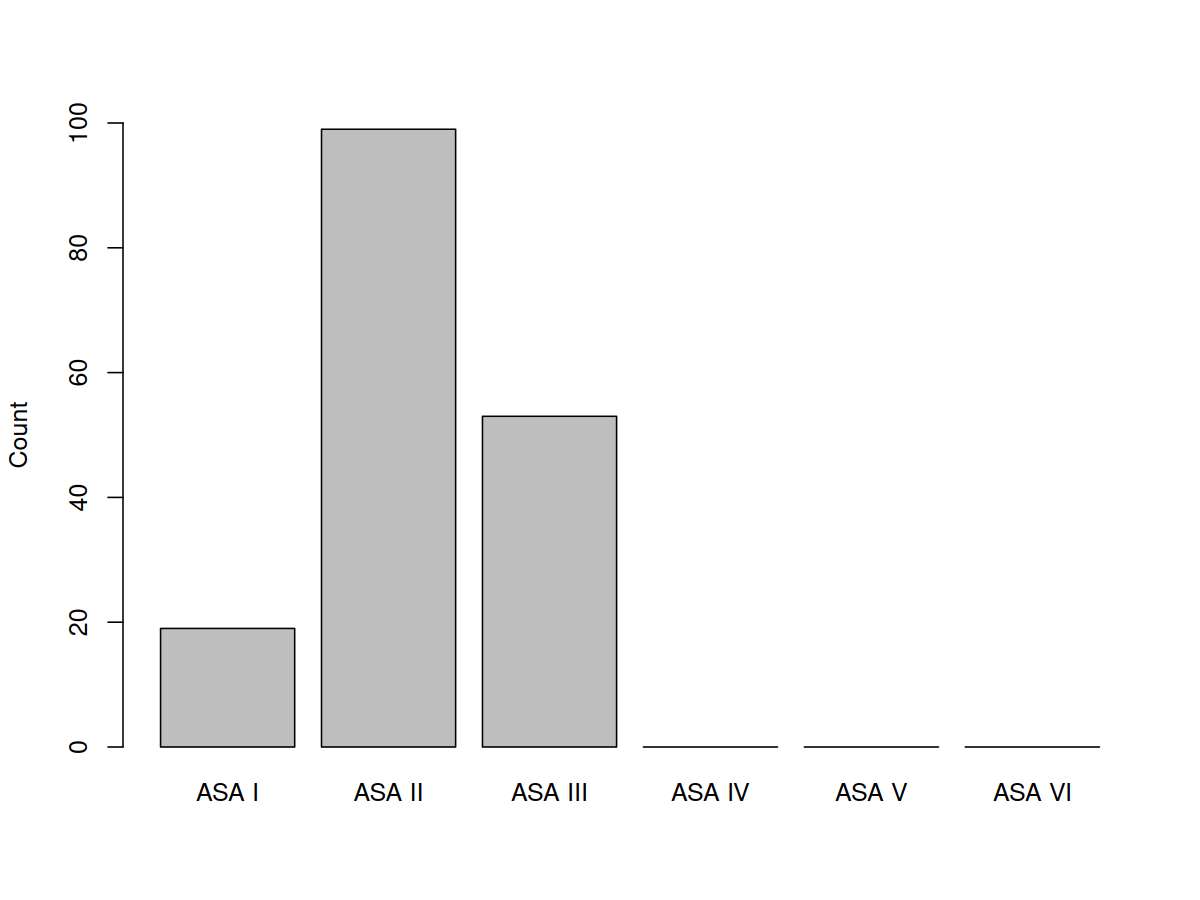}
        \caption{Physical State (ASA)}
        \label{fig:sub1}
    \end{subfigure}\\[0.7cm]
    \begin{subfigure}[b]{0.49\textwidth}
        \centering
        \includegraphics[width=.85\linewidth,trim=0 0 0 0, clip=TRUE]{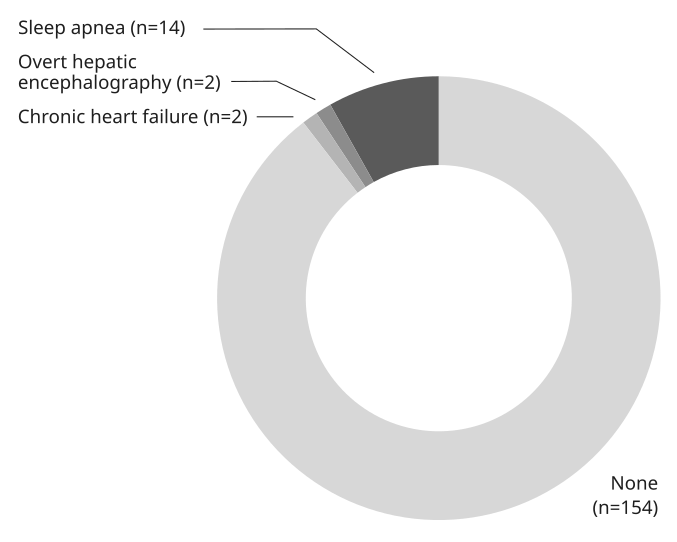}
        \caption{Preexisting Conditions}
        \label{fig:sub2}
    \end{subfigure}
    \hfill
    \begin{subfigure}[b]{0.49\textwidth}
        \centering
        \includegraphics[width=.85\linewidth,trim=0 0 0 0, clip=TRUE]{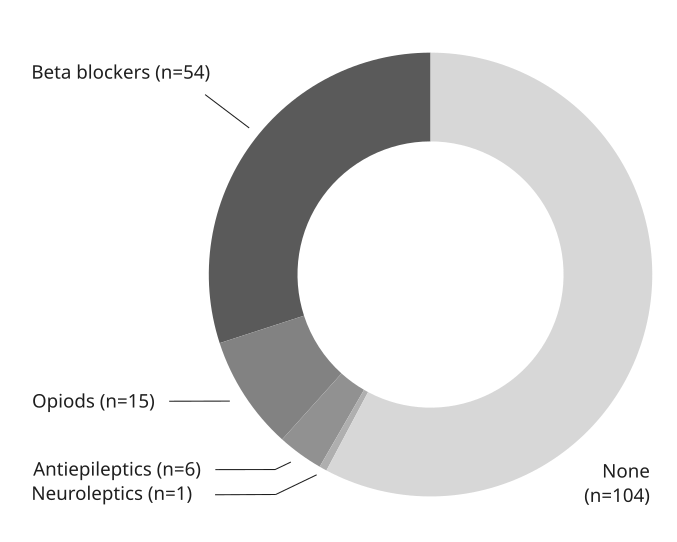}
        \caption{Relevant Prescribed Medications}
        \label{fig:sub1}
    \end{subfigure}\\[0.6cm]
    \begin{subfigure}[b]{\textwidth}
        \centering
        \includegraphics[width=.91\linewidth,trim=0 10 0 0, clip=TRUE]{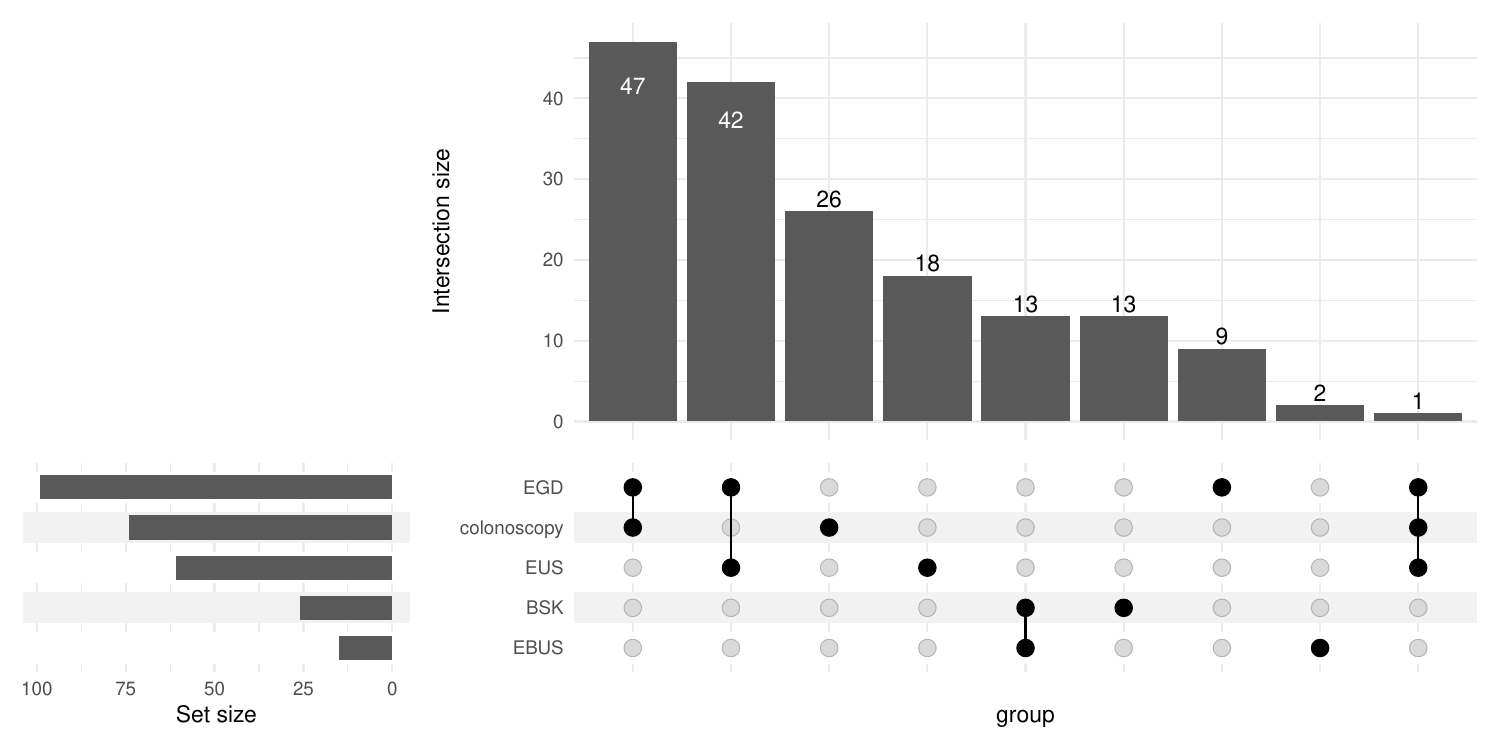}
        \caption{Type of Endoscopy}
        \label{fig:sub1}
    \end{subfigure}\\[0.1cm]
    \caption{Key Characteristics of the DOSE-I Dataset}
    \label{fig:main}
\end{figure}

Among the included recordings, 71 were collected from female patients (42\%), while 100 were collected from male patients (58\%). Patient age ranged from 21 to 83 years, with a mean age of 60.4 years, thereby covering a broad adult age spectrum typically encountered in gastrointestinal and pulmonary endoscopy units. The cohort included both younger adults undergoing elective diagnostic procedures and older patients with increased perioperative risk profiles. Figure \ref{fig:main} illustrates the distributions of key characteristics among the study cohort visually. The inclusion and exclusion criteria applied for the clinical study during patient recruitment are summarized in Table~\ref{table:inout}.

Overall, the DOSE-I dataset represents a clinically heterogeneous real-world patient population that is broadly representative of adult patients undergoing procedural sedation for endoscopic interventions in routine hospital settings. This heterogeneity is advantageous for machine learning applications because it exposes algorithms to diverse physiological patterns, patient characteristics, and sedation responses, thereby supporting the development of models with improved robustness and generalizability. 



\bigskip

\begin{table}[h!]
\centering
\small
\begin{tabularx}{\textwidth}{|X|X|}
\hline
\textbf{Inclusion Criteria} & \textbf{Exclusion Criteria} \\
\hline
\begin{itemize}[leftmargin=*]
    \item Sex: All
    \item Age: $\ge$ 18 years
    \item Ability of give informed consent
    \item Planned elective procedure in Endoscopy Department
    \item Anticipated procedure duration $\ge$ 20 min
    \item Sedation regimen: Nurse Administered Propofol Sedation
\end{itemize} 
&
\begin{itemize}[leftmargin=*]
    \item Impaired hearing
    \item Cognitive impairment (GCS $\leq$ 14) or acute psychiatric alteration
    \item Structural brain disease, cerebral metastasis, known epilepsy, or mental alteration due to new medication
    \item Sedation with substances other than those outlined in the protocol
    \item Allergy or intolerance to Propofol and its components
\end{itemize} \\
\hline
\end{tabularx}
\caption{Inclusion and Exclusion Criteria}
\label{table:inout}
\end{table}

\clearpage
\newpage

\section{Recording \& Annotation}

\subsection{Recording Setup} \label{sec:recording setup}


All physiological signals were acquired using a \textit{Philips IntelliVue MP40}, a modular clinical monitoring platform designed for continuous bedside acquisition, processing, visualization, and recording of multiple physiological signals in operating rooms and intensive care units. It supports synchronized monitoring of vital parameters, including cardiovascular and respiratory measurements, and provides real-time signal display and data integration for clinical assessment and retrospective analysis.

\begin{figure}[b!]
    \centering
    \includegraphics[width=4.1cm]{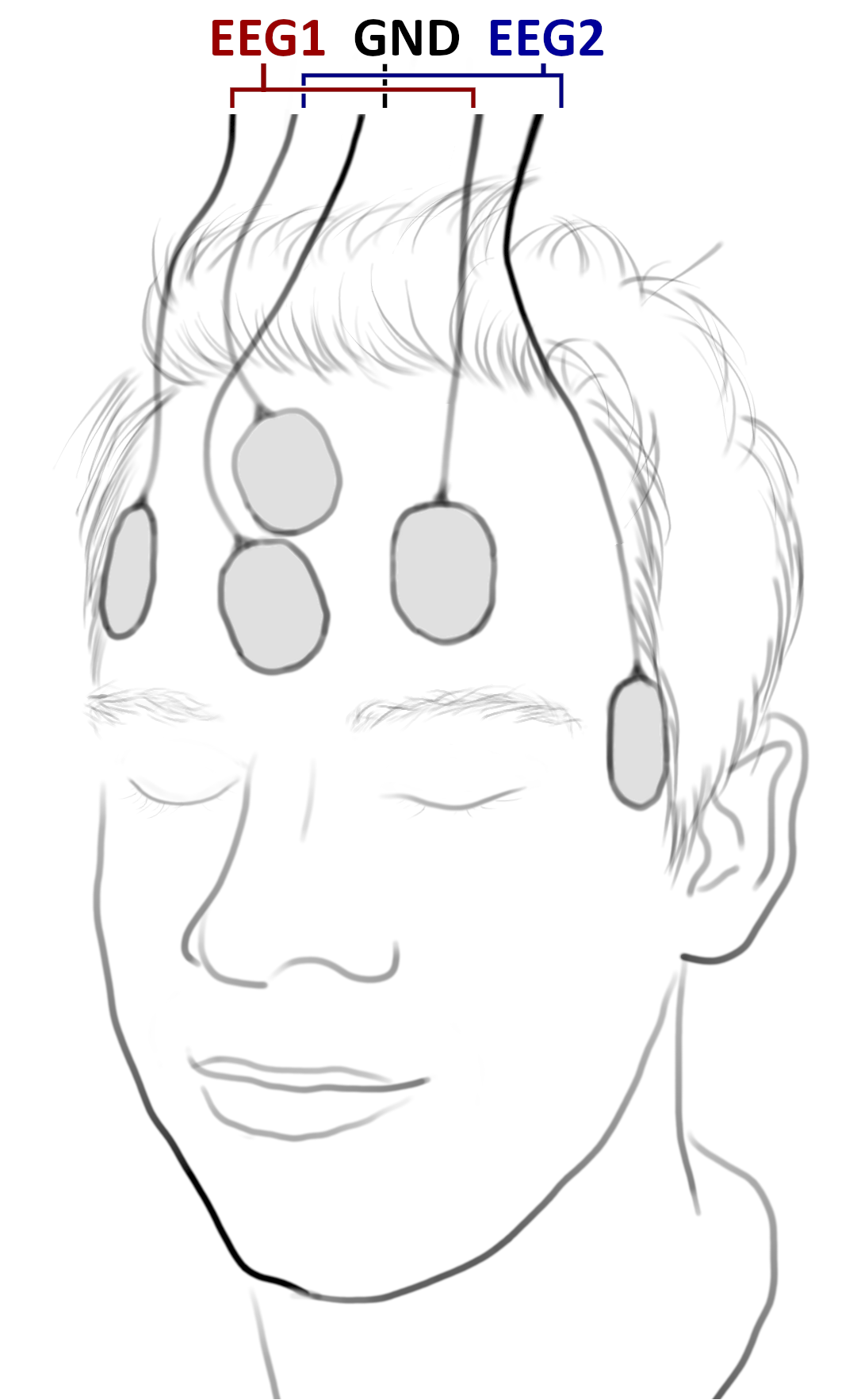}
    \caption{Electrode Configuration for Right-handed Patients.}
    \label{fig:electrodepos}
\end{figure}

Electroencephalography (EEG) was recorded using a two-channel configuration in accordance with the international 10–20 system: EEG1 (Fp2–Fp1) and EEG2 (Fz–F7), with the ground electrode positioned at Fpz (see Figure \ref{fig:electrodepos}). For left-handed participants, electrode placement was mirrored to account for hemispheric dominance. Pre-gelled Ag\/AgCl electrodes (\emph{AMBU Neuroline 720}) were used for signal acquisition. Prior to electrode placement, the  skin was cleansed and degreased with alcohol to reduce impedance. Electrode–skin impedance was maintained below 10 k$\omega$ wherever possible.

In parallel, a single-lead electrocardiogram (ECG) was recorded using limb lead II (RA–LL configuration) at a sampling frequency of 500 Hz and subsequently downsampled to 250 Hz for analysis. Photoplethysmography (PLETH) signals were obtained via a finger-mounted sensor. Non-invasive blood pressure (NIBP) measurements were performed using appropriately sized upper-arm cuffs placed contralateral to the site of Propofol administration.


To enable continuous, high-resolution data export, a MIB/RS232 interface card was installed in the patient monitor, providing a serial communication pathway for external data acquisition. The monitor was connected to a laptop computer running \textit{ixTrend Express v2.1} (ixitos GmbH, Germany), which facilitated real-time recording, visualization, and storage of all available physiological signals. It should be noted that, at the time of writing, this software is no longer commercially available or supported.

Data acquisition was initiated at least 30 seconds prior to the first administration of propofol to ensure adequate baseline recording. Recording was terminated no earlier than 30 seconds after the fulfillment of all of the following criteria: (i) completion of the endoscopic procedure, (ii) recovery of patient consciousness, and (iii) attainment of a Modified Observer’s Assessment of Alertness/Sedation (MOAA/S) score of 5. Throughout the procedure, relevant clinical events (e.g., drug administration, procedural milestones) were annotated in real time within the data stream using predefined keyboard shortcuts, enabling precise temporal alignment between physiological signals and clinical interventions.

\bigskip

\subsection{Assessment and Annotation of Consciousness States}
\emph{State of Consciousness (SOC)} and \emph{Clinical Sedation Depth} were recorded at 30-second intervals, except when safety-critical communication among the endoscopy personnel precluded interruption by the investigator. Study personnel were also permitted to perform assessments at shorter intervals if deemed necessary. \emph{SOC} was evaluated using the isolated forearm technique: the patient was instructed to squeeze the investigator's hand. If the patient did not respond, the verbal challenge was repeated after 5 seconds. If there was still no response to the second challenge, Loss of Consciousness ({\fontfamily{qcr}\selectfont LOC}) was assumed. The procedure for determining Return of Consciousness ({\fontfamily{qcr}\selectfont ROC}) was identical, but required the patient to respond adequately to both consecutive challenges. While SOC was actively tested at every timestamp where Sedation Depth markers appear, only the definitive transition points of LOC and ROC were explicitly annotated in the dataset. 

\emph{Sedation Depth} was assessed using the Modified Observer’s Assessment of Alertness/Sedation Scale (MOAA/S) by \citet{CHERNIK.1990}, a 7-point scale ranging from agitation to unresponsiveness to a painful stimulus. For the DOSE-I dataset, the extremes of this scale were grouped to make real-time data acquisition manageable for the investigators in a clinical setting. Specifically, MOAA/S scores of 5 and 6 were combined and recorded as a 5, while scores of 0 and 1 were combined and recorded as a 1.

\subsection{Annotation of Clinical/Procedural Events}
\paragraph{Endoscopy Events}{\fontfamily{qcr}\selectfont BEGIN} and {\fontfamily{qcr}\selectfont END} markers denote the insertion and removal of the endoscope, respectively. Multiple pairs of markers may be present within a single record; the most common combination represents an esophagogastroduodenoscopy (EGD) followed by a colonoscopy. The {\fontfamily{qcr}\selectfont DOSE-I\_static\_data.csv} file contains detailed information regarding the specific types of endoscopic procedures performed during each recorded session.

\paragraph{Miscellaneous}The {\fontfamily{qcr}\selectfont PARA} marker denotes the paravasation (extravasation) of propofol. Researchers should exercise caution when interpreting dose-effect relationships in these specific records, particularly in the temporal vicinity of this marker.

\subsection{Annotation of Propofol Administration}
During each procedure, administration of Propofol was documented on a continuous basis. The {\fontfamily{qcr}\selectfont PROP} marker indicates administration of propofol bolus doses in 10\,mg increments. According to the DOSE-I study protocol, no additional sedative drugs, such as midazolam, were administered during the procedure.

\subsection{Post-Processing of Annotations}
Given the dynamic clinical environment, errors in the real-time documentation of events were anticipated. Investigators were therefore provided with an {\fontfamily{qcr}\selectfont Error} marker to flag incorrectly logged events for subsequent post-hoc correction. Additionally, an automated screening process was implemented to detect implausible event sequences (e.g., mismatched endoscopy {\fontfamily{qcr}\selectfont BEGIN/END} markers or {\fontfamily{qcr}\selectfont LOC/ROC} markers). These discrepancies were corrected during post-processing whenever the ground truth could be reliably reconstructed. 

\paragraph{Note on Metadata} The {\fontfamily{qcr}\selectfont residual\_errors} column contains flagged errors from the automated detection step that could not be confidently resolved during post-processing. For further details, please refer to section \ref{sec:metadata} on Metadata.

Finally, several markers were recoded into separate, individual columns to facilitate analysis, and additional data grouping was performed specifically for the {\fontfamily{qcr}\selectfont PROP} markers. Further elaboration on this process can be found in section \ref{annotation columns} on the Annotation Columns.
\pagebreak

\section[Static Data]{Static Data\protect\footnote{Associated file: {\fontfamily{qcr}\selectfont DOSE-I\_static\_data.csv}}}


For each recording, accompanying static data were obtained from electronic case report forms (eCRFs). These data comprised variables relevant to both the pharmacokinetics and pharmacodynamics of propofol, supplemented by additional demographic and clinical descriptors to support comprehensive analysis. The dataset structure and variable definitions are summarized in Table \ref{tab:staticdatatable}.

Note that columns with unstandardized data have been redacted to ensure anonymization.

\begin{table}[h]
    \centering
   \begin{tabular}{r|c|c|l}
        Column & Unit & Dtype & Description\\
        \hline
        ID                  &           & int  & study ID \\
        ASA                 &           & int  & Am Soc of Anesth. physical status classification system\\
        age                 &           & int  & in years, rounded to nearest integer \\
        sex                 &           & string  & biological sex\\
        height              & cm        & int  & height of subject\\
        weight              & kg        & int  & weight of subject\\
        bmi                 & kg/m$^2$  & float& body mass index \\
        drugs\_bblocker     &           & boolean & use of beta blockers \\
        drugs\_opioids      &           & boolean & use of opioids\\
        drugs\_neuroleptics &           & boolean & use of neuroleptics\\
        drugs\_benzodiazepines  &       & boolean & use of benzodiazepines\\
        drugs\_antiepileptics   &       & boolean & use of antiepileptics\\
        conditions\_OHS     &           & boolean & preexisting obesity hypoventilation syndrome\\
        conditions\_SAS     &           & boolean & preexisting sleep apnea\\
        conditions\_oHE     &           & boolean & preexisting overt hepatic encephalopathy, Westhaven $\ge2$\\
        conditions\_CHF     &           & boolean & preexisting chronic heart failure, bool, NYHA III-IV\\
        handedness          &           & string  & left or right\\
        care\_type          &           & string  & care setting, in- or outpatient \\ 
        endoscopy\_type     &           & list & type of procedure performed, multiple possible \\
    \end{tabular}
    \caption{Contents of Static Data file}
    \label{tab:staticdatatable}
\end{table}

\subsection{Relevant Prescribed Medications} 
The recorded medication classes represent pharmacological groups that may influence physiological responses and the effects of {P}ropofol:
\begin{itemize}
    \item \textit{$\beta$-adrenergic blockers} (beta blockers): Reduce heart rate and blood pressure by inhibiting sympathetic activity, potentially blunting cardiovascular responses during sedation.
    \item \textit{Opioids}: Provide analgesia and sedation via $\mu$-opioid receptors; they can depress respiration and act synergistically with propofol.
    \item \textit{Neuroleptics} (antipsychotics): Primarily dopamine antagonists with sedative and autonomic effects, which may alter cardiovascular and EEG signals.
    \item \textit{Benzodiazepines}: Enhance GABA receptor activity, producing sedation and anxiolysis; they exhibit strong synergy with propofol and significantly affect EEG patterns.
    \item \textit{Antiepileptic drugs}: Stabilize neuronal excitability through various mechanisms and may modify baseline EEG activity and responses to anesthetics.
\end{itemize}

These drug classes were included to account for potential confounding effects on pharmacodynamics, pharmacokinetics, and recorded physiological signals.

\subsection{Preexisting Conditions} 
These variables represent relevant preexisting comorbidities encoded as binary indicators (present/absent): 
\begin{itemize}
    \item \textit{Obesity hypoventilation syndrome} (OHS): associated with chronic hypoventilation and increased susceptibility to oxygen desaturation and respiratory depression under sedation.
    \item \textit{Sleep apnea syndrome} (SAS): characterized by recurrent upper airway obstruction during sleep, leading to intermittent hypoxemia and increased risk of airway compromise during sedation.
    \item \textit{Overt hepatic encephalopathy} (oHE): defined as West Haven grade $\geq 2$, reflecting clinically significant hepatic dysfunction with altered cerebral function and potential baseline EEG abnormalities as well as altered response to propofol.
    \item \textit{Chronic heart failure} (CHF): defined as NYHA class III–IV, indicating markedly reduced cardiac reserve and limited functional capacity.
\end{itemize}

These conditions were included to account for clinical conditions that may influence respiratory, cardiovascular, and neurophysiological responses during {P}ropofol sedation and endoscopic procedures. They may act as confounders for both pharmacodynamic effects and physiological signal interpretation.

Data on comorbidities were obtained through clinical history and medical records; however, the latter exhibited significant gaps regarding outpatients. Researchers utilizing this dataset must account for a substantial risk of incomplete documentation, potentially leading to an underestimation of disease prevalence.

\subsection{Endoscopy Type}
The type of endoscopy conducted during a given recording is specified in the {\fontfamily{qcr}\selectfont endoscopy\_type} column as follows:
\begin{itemize}
    \item \textbf{EGD:} esophagogastroduodenoscopy, an upper gastrointestinal endoscopic procedure used for diagnostic and therapeutic evaluation of the esophagus, stomach, and duodenum.
    \item \textbf{EUS:} endoscopic ultrasound, combining endoscopy and ultrasonography for high-resolution imaging of the gastrointestinal wall and adjacent structures.
    \item \textbf{BSK:} bronchoscopy, an endoscopic technique for visualization and intervention within the tracheobronchial tree.
    \item \textbf{EBUS:} endobronchial ultrasound, an extension of bronchoscopy with ultrasound guidance for assessment and sampling of mediastinal and hilar lymph nodes.
\end{itemize}

\subsection{Other} 

The dataset includes additional patient- and procedure-related descriptors to contextualize physiological measurements and sedation responses.

The \textit{American Society of Anesthesiologists (ASA) Physical Status Classification} is recorded as an ordinal variable reflecting the patient’s preoperative health status, ranging from ASA I (healthy) to ASA V (moribund). This classification provides a standardized assessment of systemic disease burden and is commonly used to estimate perioperative risk.

\textit{Handedness} is documented as a categorical variable (left- or right-handed). This information is particularly relevant for neurophysiological measurements, as hemispheric dominance may influence EEG patterns and electrode placement considerations.

The \textit{Care Type} variable distinguishes between inpatient and outpatient cases. It is generally assumed that inpatients exhibit higher morbidity and clinical acuity.



\vfill
\pagebreak

\section{Biosignals}

\subsection[Primary Biosignals]{Primary Biosignals\protect\footnote{Associated files: {\fontfamily{qcr}\selectfont 10\_XXX.csv}, where {\fontfamily{qcr}\selectfont XXX} is zero-padded study ID.}}

Primary biosignals refer to signals directly acquired from patient monitoring devices, as opposed to derived or secondary features. The DOSE-I dataset contains five primary biosignals (cf. Tab. \ref{tab:biosignals}):
\begin{enumerate}
    \itemsep 1mm
    \item Electrocardiograms record the heart’s electrical activity from the body surface and are used to assess cardiac rhythm and function.
    \item Photoplethysmograms measure blood volume changes in peripheral tissue using optical sensing, enabling estimation of heart rate and perfusion.
    \item Electroencephalograms capture brain electrical activity via scalp electrodes, reflecting neural dynamics in the microvolt range.
    \item ECG-derived respiration estimates respiratory activity from modulation patterns in the ECG signal.
    \item Non-invasive blood pressure provides intermittent cuff-based measurements of systolic and diastolic pressure, reflecting cardiovascular state without continuous waveform recording.
\end{enumerate}

\begin{table}[h]
    \centering
    \small
    \begin{tabular}{c|c|c|c|c|c|c|c}
        Primary Biosignal & Abbreviation & sfreq nom. & sfreq & sfreq data & Range & Unit \\
        \hline
        Electrocardiogram & ECG & 500 Hz & 499.17 Hz & 250 Hz & & mV  \\
        Photoplethysmogram & PLETH & 125 Hz & 124.28 Hz & 125 Hz & &   \\
        Electroencephalogram & EEG & 125 Hz & 124.28 Hz & 125 Hz & $\pm$ 187.5 & $\mu$V \\
        ECG-derived respiration & RESP & 62.5 Hz & & 62.5 Hz & & $\Omega$ \\
        Non-invasive blood pressure & NIBP & & & & & mmHg \\
    \end{tabular}
    \caption{Description of biosignals in DOSE-I Dataset}
    \label{tab:biosignals}
\end{table}

\textbf{Sampling Rates.} Differences between the manufacturer-specified sampling frequency (nominal sampling frequency, sfreq nom.) and the empirically observed sampling frequency (sfreq) may lead to temporal misalignment. When signals are resampled to a common grid, this results in irregular samples and isolated missing values. Therefore, ECG signals were downsampled to 250 Hz to harmonize them with downstream processing requirements. PLETH and EEG signals were retained at their recorded sampling rates. 


\bigskip

\subsection{Secondary Numeric Parameters}
Four secondary signals have been derived from primary biosignals and captured at 1 Hz resolution:

\begin{enumerate}
    \item \emph{Heart Rate (HR)} derived from ECG (\texttt{ECG\_HR}) and PLETH (\texttt{PLETH\_HR})
    \item \emph{Blood Oxygen Saturation (\texttt{SAT\_O2})} derived from PLETH signal
    \item \emph{Perfusion Index (\texttt{PLETH\_PERF\_REL})} as an indicator of PLETH signal quality
    \item \emph{Respiratory Rate (\texttt{RR})} derived from RESP signal
\end{enumerate}

Further information on signal processing by the patient monitor can be found in the \href{https://www.documents.philips.com/assets/Instruction%20for%20Use/20220223/bbf4b3c2ca3c4a02b834ae45013659c0.pdf}{Philips Instructions for Use} accessible for free online.

\vfill

\subsection[Plots]{Plots\protect\footnote{Associated files: {\fontfamily{qcr}\selectfont 10-XXX\_plot.png}, where {\fontfamily{qcr}\selectfont XXX} is zero-padded study ID.}}

To facilitate intuitive visual interpretation of individual records, the dataset provides a multi-panel plot for each recording. This visualization is designed to provide a compact overview of physiological measurements, clinical annotations, administered medication, and neurophysiological signals over time. Fig.~\ref{fig:plot} displays an example plot for a single recording. 

The top panel ({\fontfamily{qcr}\selectfont VITALS}) displays temporal trends of key physiological measurements routinely monitored during the procedure. These include heart rate ({\fontfamily{qcr}\selectfont HR}), peripheral blood oxygen saturation ({\fontfamily{qcr}\selectfont SpO2}), and non-invasive blood pressure measurements, separated into systolic ({\fontfamily{qcr}\selectfont NIBP SYS}) and diastolic ({\fontfamily{qcr}\selectfont NIBP DIA}) values. Displaying these parameters together enables assessment of the patient’s physiological state and its evolution throughout the recording.

Clinical state and events are shown in the second panel ({\fontfamily{qcr}\selectfont REF}), which includes episodes of unconsciousness and corresponding MOAA/S scores. Drug administration is visualized in the {\fontfamily{qcr}\selectfont DRUGS} panel, where propofol boluses are indicated. Additional procedural and miscellaneous annotations, such as endoscopy events and other unclassified markers (e.g., {\fontfamily{qcr}\selectfont PARA}), are collected in the {\fontfamily{qcr}\selectfont OTHER} panel.


Neurophysiological recordings acquired during the endoscopy procedure are displayed in panels {\fontfamily{qcr}\selectfont EEG1} and {\fontfamily{qcr}\selectfont EEG2}, which reflect individual EEG channels. EEG segments identified as artifacts by an automated detection algorithm are highlighted in red. These marked segments typically correspond to portions of the signal affected by motion, noise, or other recording disturbances that may reduce signal quality and complicate interpretation.

\bigskip

\begin{figure}[h!]
    \centering
    \includegraphics[width=\textwidth, trim=10 10 10 0, clip=true]{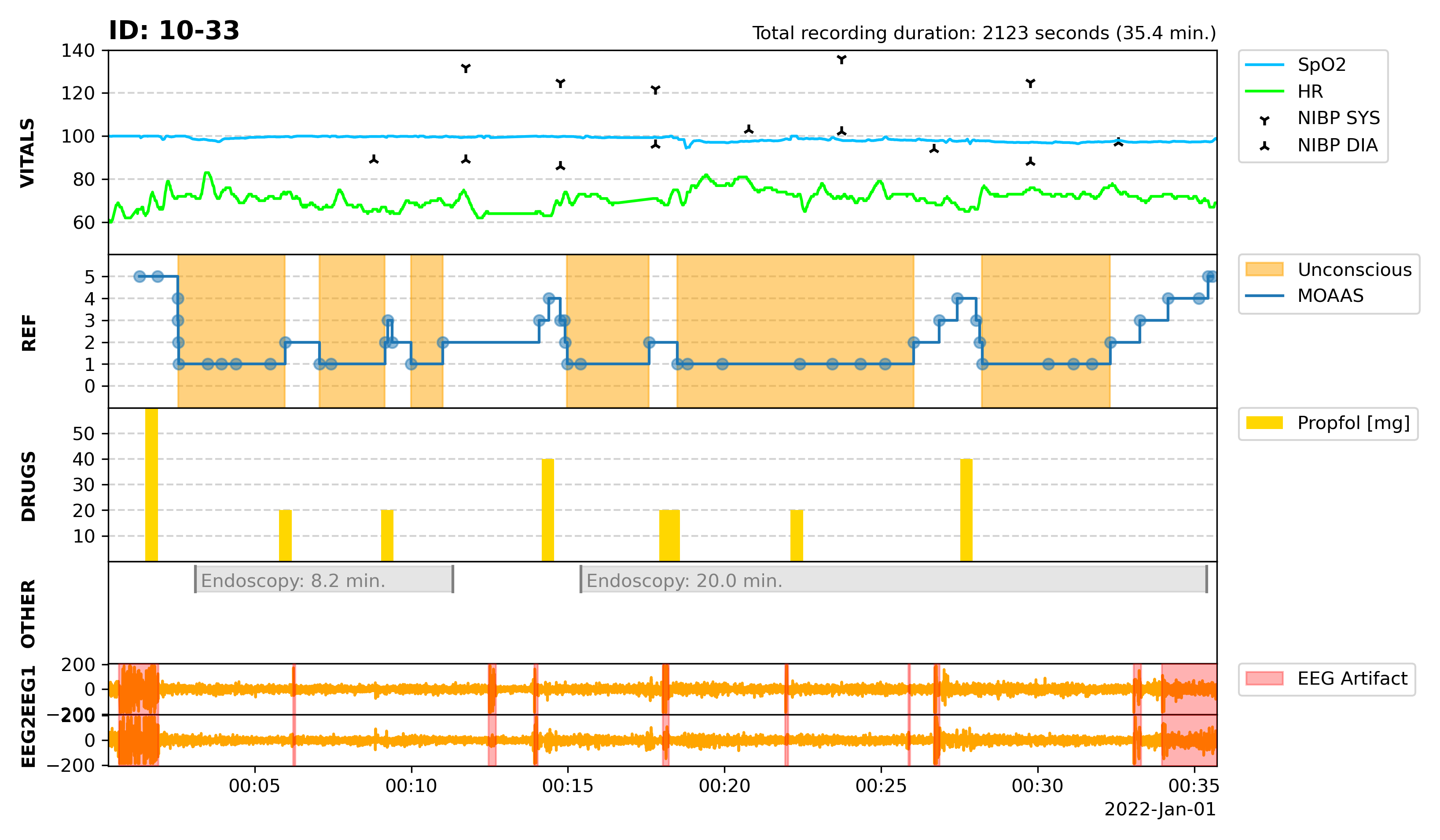}
    \caption[Sample Plot from DOSE-I Dataset Detailing Multiple Aspects of the Data Captured]{A sample plot from DOSE-I data detailing many aspects of the data captured. }
    \label{fig:plot}
\end{figure}

\vfill
\pagebreak

\subsection[Processed EEG]{Processed EEG\protect\footnote{Associated files: {\fontfamily{qcr}\selectfont 10-XXX\_pEEG.csv}, where {\fontfamily{qcr}\selectfont XXX} is zero-padded study ID.}}

In addition to raw EEG data, a set of 40 processed EEG (short: pEEG) parameters  are provided (cf. Table \ref{tab:paramslit}).  Derived from raw EEG signals, these numerical features simplify brain activity into values clinicians can use—especially to monitor depth of anesthesia or sedation. 

\begin{table}[!h]
    \centering
    \small
    \begin{tabular}{c|c|c|c}
        Column & Col Name & Frequency Range and Parameters & Reference \\
        \hline
        1 & time & & \\
        2 & abs\_subdelta & 0.5$-$1 Hz & \\
        3 & abs\_delta1 & 1$-$2 Hz & \\
        4 & abs\_delta2 & 2$-$4 Hz & \\
        5 & abs\_theta & 4$-$7.5 Hz & \\
        6 & abs\_alpha & 7.5$-$13 Hz & \\
        7 & abs\_beta1 & 13$-$20 Hz & \\
        8 & abs\_beta2 & 20$-$30 Hz & \\
        9 & abs\_gamma & 30$-$49 Hz & \\
        10 & rel\_subdelta & 0.5$-$1 Hz & \\
        11 & rel\_delta1 & 1$-$2 Hz & \\
        12 & rel\_delta2 & 2$-$4 Hz & \\
        13 & rel\_theta & 4$-$7.5 Hz & \\
        14 & rel\_alpha & 7.5$-$13 Hz & \\
        15 & rel\_beta1 & 13$-$20 Hz & \\
        16 & rel\_beta2 & 20$-$30 Hz & \\
        17 & rel\_gamma & 30$-$49 Hz & \\
        18 & sync\_subdelta & 0.5$-$1 Hz & \citet{rosenblum2001phase} \\
        19 & sync\_delta1 & 1$-$2 Hz & \citet{rosenblum2001phase} \\
        20 & sync\_delta2 & 2$-$4 Hz & \citet{rosenblum2001phase} \\
        21 & sync\_theta & 4$-$7.5 Hz & \citet{rosenblum2001phase} \\
        22 & sync\_alpha & 7.5$-$13 Hz & \citet{rosenblum2001phase} \\
        23 & sync\_beta1 & 13$-$20 Hz & \citet{rosenblum2001phase} \\
        24 & sync\_beta2 & 20$-$30 Hz & \citet{rosenblum2001phase} \\
        25 & sync\_gamma & 30$-$49 Hz & \citet{rosenblum2001phase} \\
        26 & MF & 0.5$-$30 Hz, p=2, r=0.5, T=8 s & \citet{jordan2007median} \\
        27 & SEF95 & 0.5$-$30 Hz, p=2, r=0.95, T=8 s & \citet{jordan2007median} \\
        28 & WSMF30 & 8$-$30 Hz, p=0.4, r=0.5, T=8 s & \citet{jordan2007median} \\
        29 & WSMF49 & 8$-$49 Hz, p=1.0, r=0.5, T=8 s & \citet{jordan2007median} \\
        30 & PE31 & 0.5$-$45 Hz, n=3, tau=1, tie=0.5 $\mu$V & \citet{olofsen2008permutation} \\
        31 & PE32 & 0.5$-$45 Hz, n=3, tau=2, tie=0.5 $\mu$V & \citet{olofsen2008permutation} \\ 
        32 & PE61 & 0.5$-$45 Hz, n=6, tau=1, no tie & \citet{jordan2008electroencephalographic} \\
        33 & SFS & 40$-$47 Hz over 1$-$47 Hz & \citet{miller2004bispectral} \\
        34 & CFS & 40$-$47 Hz over 1$-$47 Hz & \citet{rampil1998primer} \\
        35 & PFS & 40$-$47 Hz over 1$-$47 Hz & \citet{miller2004bispectral} \\
        36 & WSMF\_Klimpel & 8.5$-$27.5 Hz, p=0.41, r=0.329, T=8 s & \citet{klimpel2020eeg} \\
        37 & MF\_Jordan & 0.5$-$30 Hz, p=2, r=0.5, T=16 s & \citet{jordan2007median} \\
        38 & SEF95\_Jordan & 0.5$-$30 Hz, p=2, r=0.95, T=16 s & \citet{jordan2007median} \\ 
        39 & WSMF30\_16 & 8$-$30 Hz, p=0.4, r=0.5, T=16 s & \citet{jordan2007median} \\
        40 & WSMF49\_16 & 8$-$49 Hz, p=1.0, r=0.5, T=16 s & \citet{jordan2007median} \\
        41 & WSMF\_Klimpel\_16 & 8.5$-$27.5 Hz, p=0.41, r=0.329, T=16 s & \citet{klimpel2020eeg} \\
        \hline
        \multicolumn{4}{l}{\footnotesize MF: Median Freq., SEF95: Spectral Edge Freq. 95\%, WSMF: Weighted Spectral Median Freq.,}\\
        \multicolumn{4}{l}{\footnotesize PE: Permutation Entropy, SFS: SynchFastSlow, CFS: Bicoherence of SFS, PFS: PowerFastSlow}\\
        \multicolumn{4}{l}{\footnotesize Note: CPEI is defined as the average of the two PE measures according to Olofsen.}\\
    \end{tabular}
    \caption{Overview of all 40 pEEG parameters (18 to 41 taken from the literature).}
    \label{tab:paramslit}
\end{table} 

 All parameters were calculated at a rate of one value per second. The first column {\fontfamily{qcr}\selectfont Time} of each record holds a timestamp with second-level precision detailed in section \ref{parsing}. For convenience in loading the data, the last columns ({\fontfamily{qcr}\selectfont 42} to {\fontfamily{qcr}\selectfont 49}) are the numeric biosignals (HR, RR, SAT\_O2) and annotations columns from the data files (see section \ref{annotation columns}) with a notable difference: Columns 47 (MOAAS), 48 (SOC) and 49 (Endoscopy) were forward filled. 

\subsubsection*{pEEG Parameters Derived from EEG Bands}

The first three types of parameters were derived for each of eight considered EEG bands, see Tab. \ref{tab:bandparams}. To split the EEGs into these bands, FFT windows with lengths corresponding to an average of 10 oscillations were used for each band. These windows overlapped by 50\%. Specifically, 
\begin{itemize}
    \item the logarithm of the \emph{absolute} EEG power in each band, e.g., abs\_delta$0=\log_{10}$(power$(0.5 - 1$ Hz))
    \item the logarithm of the \emph{relative} EEG power (absolute power divided by the sum of powers for all bands), e.g., rel\_delta$0=\log_{10}$(power$(0.5 - 1$ Hz) / power$(0.5 - 49$ Hz)), and
    \item \emph{phase synchronisation} between the two EEG signals
\end{itemize}
were analysed. For the phase synchronization calculations we refer to Rosenblum et al. (2001).
 
\begin{table}[h]
    \centering
    \small
    \begin{tabular}{c|c|c|c|c}
        \multicolumn{1}{c|}{frequency band} & \multicolumn{1}{c|}{range [Hz]} &\multicolumn{3}{c}{Column in files} \\
        \cline{3-5}
        & & log\_(abs. Pwr) & log\_(rel. Pwr) & Phase Synchron. \\
        \hline 
        delta0 & 0.5$-$1 & 2 & 10 & 18 \\
        delta1 & 1$-$2 & 3 & 11 & 19 \\
        delta2 & 2$-$4 & 4 & 12 & 20 \\
        theta & 4$-$7.5 & 5 & 13 & 21 \\
        alpha & 7.5$-$13 & 6 & 14 & 22 \\
        beta1 & 13$-$20 & 7 & 15 & 23 \\
        beta2 & 20$-$30 & 8 & 16 & 24 \\
        gamma & 30$-$49 & 9 & 17 & 25 \\
    \end{tabular}
    \caption{Frequency Ranges and Columns of Band pEEG Parameters}
    \label{tab:bandparams}
\end{table}

Column names in columns 2 - 25 are constructed with a prefixes {\fontfamily{qcr}\selectfont abs\_}, {\fontfamily{qcr}\selectfont rel\_} and {\fontfamily{qcr}\selectfont sync\_} followed by the frequency band (e.g. {\fontfamily{qcr}\selectfont abs\_delta2}, {\fontfamily{qcr}\selectfont rel\_gamma} or {\fontfamily{qcr}\selectfont sync\_beta1}). \\

\subsubsection*{pEEG Parameters from the Literature}

In addition to the previously described parameters, several parameters described in literature were calculated, detailed in Table \ref{tab:paramslit} and Garbe et al. (2021). Note that CPEI is defined as the average of the two PE measures according to Olofsen et al. (2008), which are given in Columns 30 and 31. The Permutation Entropy (PE) according to Jordan et al. (2008) in Column 32 is adapted to the sampling frequency of this dataset and aligned with the previous two PE measures.

Specifically, SFS, CFS, and PFS were defined as follows:
{\small
$$ {\rm SFS} = \log_{10} \left(\frac{{\rm bispectral power}(40-47 \,{\rm Hz})}{{\rm bispectral power}(1-47 \,{\rm Hz})}\right), $$
$$ {\rm CFS} = \log_{10} \left(\frac{{\rm bicoherence}(40-47 \,{\rm Hz})}{{\rm bicoherence}(1-47 \,{\rm Hz})}\right), $$
$$ {\rm PFS} = \log_{10} \left(\frac{{\rm power}(40-47 \,{\rm Hz})}{{\rm power}(1-47 \,{\rm Hz})}\right). $$
}

\clearpage

\section[Metadata]{Metadata\protect\footnote{Associated file: {\fontfamily{qcr}\selectfont DOSE-I\_metadata.csv} }}
\label{sec:metadata}

For each recording, metadata fields were recorded in collecting the DOSE-I dataset. A tabular description of columns in the metadata file is given in Tab. \ref{tab:metadatatable}.

\begin{table}[h!]
    \centering
    \begin{tabular}{r|c|c|l}
        column      & unit  & dtype   & description and further information \\
        \hline
        ID          &       & int   & study ID, corresponds to 10-X data files \\
        version     &       & string   & post-processing software version \\
        length      & s     & int   & length of record truncated according to protocol \\
        length\_vital & s   & int   & length of record before processing \\
        endoscopies &       & int   & number of endoscopies \\
        SOC\_trans  &       & int   & SOC transitions, both LOC and ROC are counted \\
        MOAAS\_sum  &       & int   & number of MOAA/S records \\
        MOAAS1      &       & int   & number of combined MOAA/S 
        Anpassung Jan p=1/0.4 @research
Anpassung Jan Formel @research
Anpassung Jakob metadata @research
Anpassung Jakob tail/end @research0 and MOAA/S 1 records \\
        MOAAS2      &       & int   & number of MOAA/S 2 records\\
        MOAAS3      &       & int   & number of MOAA/S 3 records\\
        MOAAS4      &       & int   & number of MOAA/S 4 records\\
        MOAAS5      &       & int   & number of MOAA/S 5 records\\
        dur\_SOC0   & s     & int   & duration of unconsciousness \\
        other\_events &     & list  & list of events not covered elsewere \\
        PROP\_sum   & mg    & int   & cumulative amount of Propofol given \\
        wave\_tracks &      & list  & available wave tracks in data, sorted \\
        numerics\_tracks &  & list  & available numerics tracks in data, sorted \\
        compl\_ecg\_base &  & fraction & completeness of ECG signal \\
        compl\_eeg1 &       & fraction & completeness of EEG 1 \\
        compl\_eeg2 &       & fraction & completeness of EEG 2 \\
        compl\_pleth &      & fraction & relative completeness of PLETH signal \\
        NIBP\_sum   &       & int   & number of NIBP readings \\
        residual\_errors &  & list  & errors remaining after post-processing\\
    \end{tabular}
    \caption{Description of Metadata contents}
    \label{tab:metadatatable}
\end{table}

\subsection{Consciousness}

Leveraging the so-called MOAA/S scale, the patient’s sedation depth is approximated. The variables MOAA/S1 to MOAA/S5 capture the distribution of observed sedation levels according to the Modified Observer’s Assessment of Alertness/Sedation (MOAA/S) scale by \citet{CHERNIK.1990}. Each variable represents the number of assessments falling within a specific score category, ranging from deep sedation (MOAA/S 0–1) through intermediate levels of sedation (MOAA/S 2-3) to reduced and full alertness (MOAA/S 4-5). The distribution of MOAA/S scores over time (MOAA/S1–5) captures the depth and stability of sedation, while MOAA/S\_sum reflects the temporal sampling density of clinical assessments.



Both Loss of Consciousness (LOC) and Return of Consciousness (ROC) events count towards State-of-Consciousness transitions (SOC\_trans). A high count may be reflective of high sedation depth dynamic during the procedure. The duration of unconsciousness (dur\_SOC0) quantifies the cumulative time spent in the unconsciousness. Together, these metrics offer a structured and temporally resolved representation of consciousness as inferred from observable behavioral responsiveness.

\begin{figure}[h!]
    \centering
    \includegraphics[width=.9\textwidth, trim=0 35 0 55, clip=true]{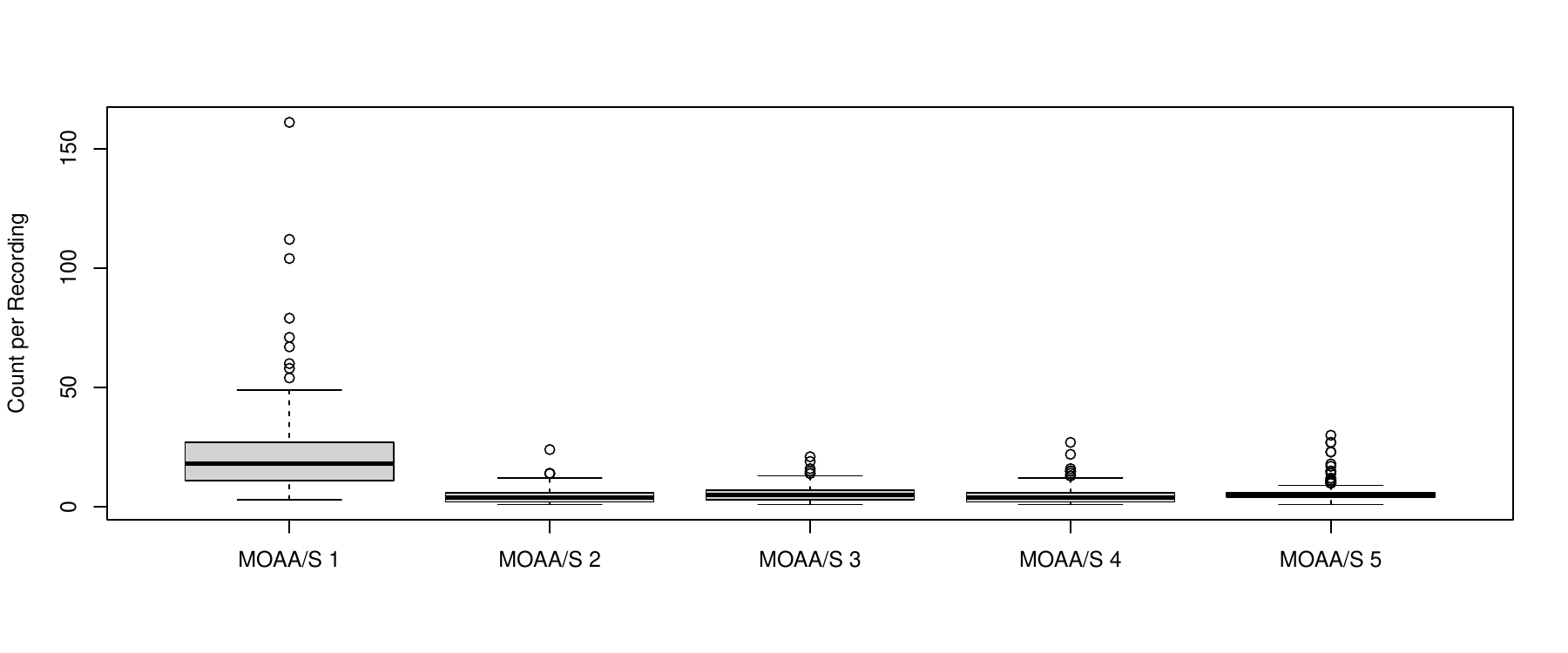}
    \caption{Boxplot showing the frequency of MOAA/S states aggregated for all 171 recordings.}
    \label{fig:plot}
\end{figure}

\subsection{Completeness}

The parameters \texttt{wave\_tracks} and \texttt{numerics\_tracks} describe the availability of recorded data channels, provided as sorted lists of waveform signals (e.g., ECG, EEG, PLETH) and numerical variables (e.g., heart rate, blood pressure), respectively. 

Signal completeness is quantified for each modality as the proportion of valid, non-missing samples relative to the expected recording duration, expressed as a fraction. Specifically, \texttt{compl\_ecg\_base}, \texttt{compl\_eeg1}, \texttt{compl\_eeg2}, and \texttt{compl\_pleth} represent the completeness of the ECG, EEG channel 1, EEG channel 2, and photoplethysmography signals. 

In addition, \texttt{NIBP\_sum} denotes the total number of non-invasive blood pressure measurements acquired during the recording period. 

Together, these parameters provide a quantitative assessment of data availability for subsequent analysis. Signal quality is discussed in appendix section B.3.

\begin{figure}[h!]
    \centering
    \includegraphics[width=.88\textwidth, trim=0 35 0 50, clip=true]{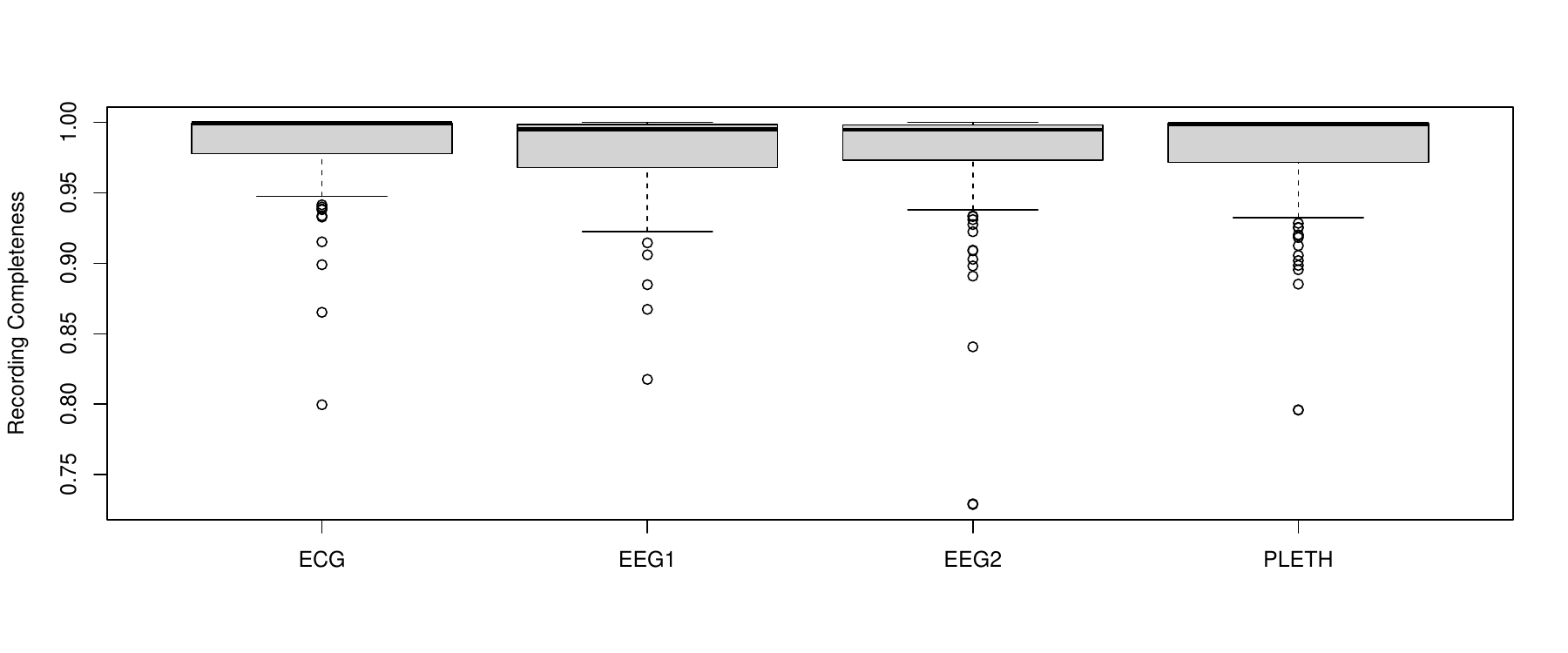}
    \caption{Boxplot showing the relative completeness of ECG, EEG and photoplethysmography signals over all 171 recordings (1 equals 100 percent complete).}
    \label{fig:plot}
\end{figure}

\subsection{Other}

The {\fontfamily{qcr}\selectfont residual\_error} columns contains three types of errors:
\begin{description}[align=right,labelwidth=3cm]
    \item[Lead] less than 30s before first dose of Propofol
    \item[Trail] less than 30s after END, MOAA/S5 and ROC
    \item[Completeness] END, MOAA/S5 or ROC not reached at end of record
\end{description}

\vfill
\pagebreak

\appendix
\section*{Appendix}
\addcontentsline{toc}{section}{Appendix}
\clearpage

\section{Data Handling}

\subsection{Structure of Data Files}
The data is anchored to a single, continuous timestamp column ({\fontfamily{qcr}\selectfont Time}) with millisecond precision. The base temporal resolution of the file is 250 Hz. This is followed by waveform and then numeric biosignal columns and lastly clinical annotation columns.

All signals have a prefix from the originating device. For DOSE-I data this is only {\fontfamily{qcr}\selectfont Intellivue/}. There are 5 waveform signals:

{\fontfamily{qcr}\selectfont
ECG\_II, EEG\_1, EEG\_2, PLETH, RESP
} 

\noindent and 8 numeric signals: 

{\fontfamily{qcr}\selectfont
ECG\_HR, RR, PLETH\_SAT\_O2, PLETH\_HR, PLETH\_PERF\_REL, \\
\indent NIBP\_SYS, NIBP\_DIA, NIBP\_MEAN
}

\paragraph{Annotation columns} \label{annotation columns}
The dataset contains seven columns dedicated to clinical annotations. The primary {\fontfamily{qcr}\selectfont EVENT} column serves as the master record, containing all annotations following error correction. To facilitate streamlined data analysis, the subsequent six columns represent specifically coded excerpts extracted directly from the {\fontfamily{qcr}\selectfont EVENT} column: 
\begin{itemize}
    \item {\fontfamily{qcr}\selectfont Misc} Primarily for future data collection compatibility. It contains any markers not explicitly recoded into the other dedicated columns. In the DOSE-I dataset, this column exclusively contains the {\fontfamily{qcr}\selectfont PARA} (paravasation) marker, which only appears in records 106 and 165.
    \item {\fontfamily{qcr}\selectfont SOC} State of Consciousness transitions. Loss of Consciousness (LOC) is coded as 0, and Return of Consciousness (ROC) is coded as 1.
    \item {\fontfamily{qcr}\selectfont MOAAS} Sedation depth scores based on the MOAA/S scale. As noted previously, scores of 0 and 1 are combined and coded as 1, while scores of 5 and 6 are combined and coded as 5.
    \item {\fontfamily{qcr}\selectfont Propofol} Administered boluses of propofol, recorded in 10 mg increments (e.g. $30 \equiv 30$mg)
    \item {\fontfamily{qcr}\selectfont Midazolam} Administered boluses of midazolam, recorded in mg. This column exists solely for structural compatibility with future data collection and remains entirely empty throughout the DOSE-I data.
    \item {\fontfamily{qcr}\selectfont Endoscopy} Endoscope insertion and removal events. {\fontfamily{qcr}\selectfont BEGIN} coded as 1, and {\fontfamily{qcr}\selectfont END} coded as 0.
\end{itemize}

\subsection{Parsing Timestamps} \label{parsing}
Each file contains a timestamp column called \emph{Time}. The date (1 Jan 2022), as well as the time, has no semantic meaning. For reasons of compatibility with future data collection, the timestamp was reformatted. Format: \\

{\fontfamily{qcr}\selectfont
YYYY-MM-DD HH:mm:ss.fff
} \\

The timestamp is \emph{free of duplicate values}, \emph{monotonically increasing}, and has \emph{fixed 4 ms intervals}, e.g.: \\

{\fontfamily{qcr}\selectfont
2022-01-01 00:19:35.436\\
\indent2022-01-01 00:19:35.440\\
\indent2022-01-01 00:19:35.444\\
\indent2022-01-01 00:19:35.448\\
} \\

\pagebreak

\section[Data Quality]{Data Quality\protect\footnote{Associated files: {\fontfamily{qcr}\selectfont 10-XXX\_artifacts.csv}, where {\fontfamily{qcr}\selectfont XXX} is zero-padded study ID.}}

\subsection{Impact of Environment and Hardware Constraints}
Unlike the highly controlled environments of the intensive care unit (ICU) or general anesthesia — where the use of muscle relaxants minimizes patient movement — endoscopy presents a much more dynamic clinical setting. Consequently, researchers utilizing this dataset should expect higher rates of electrode disconnects and movement-induced signal artifacts. This reality poses a specific challenge for electroencephalography (EEG) preprocessing. While many standard EEG artifact detection algorithms rely on spatial context and cross-channel comparisons to identify single-channel noise, these methods are incompatible with our focused two-channel setup. To address these unique environmental and hardware constraints, we have implemented a simplified, custom artifact detection pipeline to ensure transparency and baseline data reliability for future analyses.

\subsection{Signal Completeness}
The DOSE-I dataset contains 5 waveform biosignal data channels: single-lead electrocardiogram (ECG), plethysmogram (PLETH), frontotemporal 2-channel electroencephalogram (EEG1 \& EEG2), and ECG-derived respiration (RESP). Channels were recorded at 500, 125 and 62.5Hz, respectively. In post-processing the ECG channel was downsampled to 250 Hz. Missing data percentages in these channels are listed in Tab. 1.
\begin{table}[h]
    \centering
    \begin{tabular}{c|c|c|c}
        biosignal & mean [\%] & median [\%] & range\\
        \hline
        ECG & 1.4 & 0.1 & 0 - 20.0 \\
        PLETH & 2.0 & 0.1 & 0 - 20.4 \\
        EEG Ch1 & 1.9 & 0.5 & 0 - 18.2 \\
        EEG Ch2 & 2.1 & 0.5 & 0 - 27.1 \\
        RESP & 4.1 & 2.0 & 0 - 69.3\\
    \end{tabular}
    \caption{Missing data in waveform biosignals}
    \label{tab:my_label}
\end{table}

\subsection{Signal Quality}
For the ECG, EEG and RESP channels the monitor used - a \emph{PHILIPS IntelliVue MP40} - did not provide an extractable signal quality measure. For EEG channels an impedance warning was given whenever a predetermined threshold was exceeded. However, these technical warning events could not be extracted.

\paragraph{PLETH signal quality}
For the PLETH signal \emph{PLETH\_PERF\_REL} gives a value for the pulsatile portion of the measured signal caused by the pulsating arterial blood flow. According to the manufacturer, values above 1 are optimal, between 0.3 - 1 is acceptable and values below 0.3 should be interpreted with care. The value is output in 1-second intervals. 16,850 out of 268,405 readings (6.2\%) were below 1, with a range of 0 - 67.7\% and median of 0.7\%.

\subsection{EEG Signal Quality and Artifacts}
\paragraph{Rationale}
Standardized EEG preprocessing pipelines inherently rely on high-density, multiple-channel arrays to spatially isolate and regress artifacts using techniques like Independent Component Analysis (ICA). Furthermore, conventional artifact rejection methods are designed to discard physiological phenomena such as muscle activity (EMG) or eye blinks (EOG). In the context of our dataset, however, these phenomena are not noise; they hold highly relevant clinical information and serve as surrogate markers for tracking states of consciousness and sedation depth. Because our objective requires preserving this physiological information within a two-channel setup, we implemented a simple custom, highly targeted artifact detection pipeline.

\paragraph{Methodology}
The continuous EEG signal was processed through a sequence of chunk-wise evaluations and post-processing rules:
\begin{enumerate}
    \item \emph{Epoching}: The data was segmented into 2-second, non-overlapping epochs.
    \item \emph{Epoch-Wise Artifact Detection}: Each epoch was evaluated independently for three artifact types:
    \begin{enumerate}
        \item \emph{Missing Data}: Flagged if $>5\%$ of data points (equivalent to 13 values at a 125 Hz sampling rate) were missing. Reasoning: A small number of missing values can be reliably interpolated without negatively impacting subsequent Fast Fourier Transform (FFT) analysis.
        \item \emph{Extreme Values / Signal Saturation}: Flagged if at least one recorded value within the epoch reached the boundary of the hardware measuring range ($\pm187.5 \mu$V). Reasoning: Unlike missing data, even a single extreme saturated value will severely distort the frequency spectrum during FFT, making rejection mandatory.
        \item \emph{Loose Channel / Low-Variance Noise}: Identifying a loose electrode is challenging without hardware impedance data or a multi-channel reference. While alternative metrics were considered (e.g., relative spectral power around grid frequency, spectral flatness, or Hjorth parameters), they all require informed definitions of cutoffs. We opted for signal variance as a simpler alternative. To define an empirical cutoff, three neurologists experienced in visual EEG segmentation reviewed a 10-hour random subsample. Based on their consensus, a threshold of signal variance $<50 \mu$V$^2$ maintained for at least 5 consecutive epochs (10 seconds) provided a good trade-off. Crucially, this duration threshold prevents the misclassification of true physiological EEG suppressions (which can occur during deep sedation) as loose channel artifacts.
    \end{enumerate}
    \item \emph{Post-Processing}:
    \begin{enumerate}
        \item \emph{Dilation}: To ensure a clean signal boundary, one additional epoch immediately following any detected artifact epoch was also marked as an artifact.
        \item \emph{Minimum Segment Length}: Continuous "good" sections shorter than 4 epochs (8 seconds) were rejected, as this is the minimum window length required for the reliable extraction of most FFT-based frequency spectrum parameters.
        \item \emph{Channel\,Coupling}:\,If\,an\,artifact\,was\,flagged\,in\,one\,channel, the corresponding epoch was re\-jected in both channels to maintain strict temporal alignment for downstream bilateral analysis.
    \end{enumerate}
\end{enumerate} 
\paragraph{Descriptive Statistics}
Descriptive statistics regarding the frequency and distribution of these artifacts are detailed below. Out of 141,585 epochs 25,997 (18.4\%) were classified as artifacts, ranging from 2\% to 57\% on a record level. 4,990 epochs were classfified as missing data (3.5\%), 10,422 as saturation (7.4\%), 2,582 as loose channel (1.8\%). Post-processing added 5,572 (3.9\%) for dilation and 2,431 (1.7\%) for short useful EEG segments too short for meaningful analysis.

The exact artifact epoch markers and complete results can be found in the \emph{artifacts} folder of the dataset. The numbering in the \emph{reason} column is as follows:
\begin{enumerate}
    \itemsep0em
    \item Missing Data
    \item Saturation / Extreme values
    \item Loose Channel / Noise
    \item Dilation
    \item short "good" segments $< 8s$
\end{enumerate}

\paragraph{Clinical Observations}
When analyzing the distribution of these artifacts, several general observations emerged that directly reflect the realities of the clinical workflow during endoscopy:
\begin{enumerate}
    \item \emph{Signal Saturation} predominantly occurs at the very beginning of the procedure when the patient is still awake, moving, and being positioned on the table.
    \item \emph{Missing Data} often clustered in the center of the recordings. This typically corresponds to the moment the monitor is briefly disconnected so clinical staff can physically turn the patient's bed during the transition between an esophagogastroduodenoscopy (EGD) and a colonoscopy.
    \item \emph{Loose Channel} is a generally rare artifact as investigators monitored the technical warnings during the procedure; however, when it does occur, it tends to affect longer, contiguous sections of the recording.
\end{enumerate}

\begin{figure}[h]
    \centering
    \includegraphics[width=\textwidth, trim=50 25 20 15, clip=TRUE]{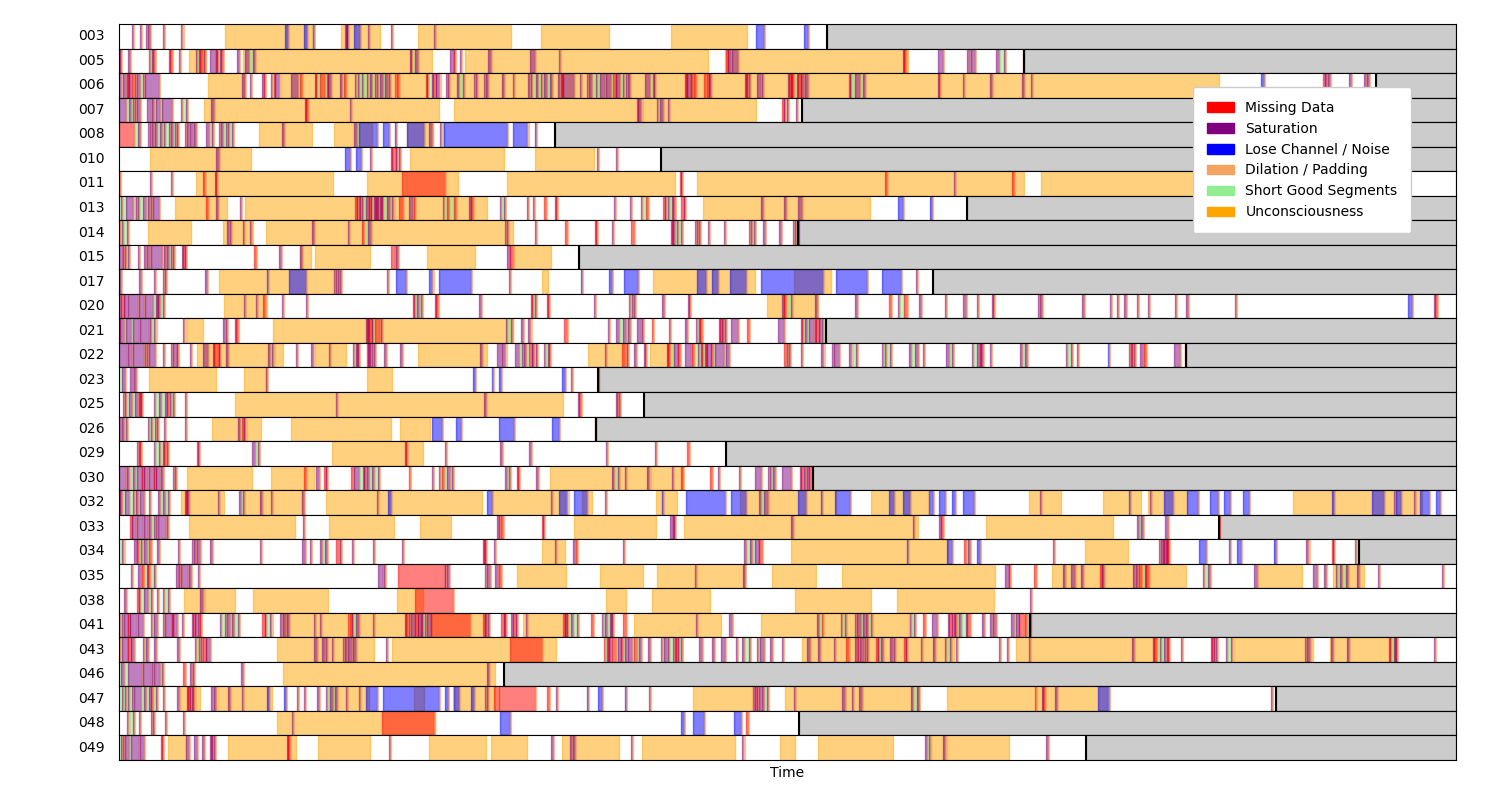}\\[-0.1cm]
    \includegraphics[width=.995\textwidth, trim=50 25 20 15, clip=TRUE]{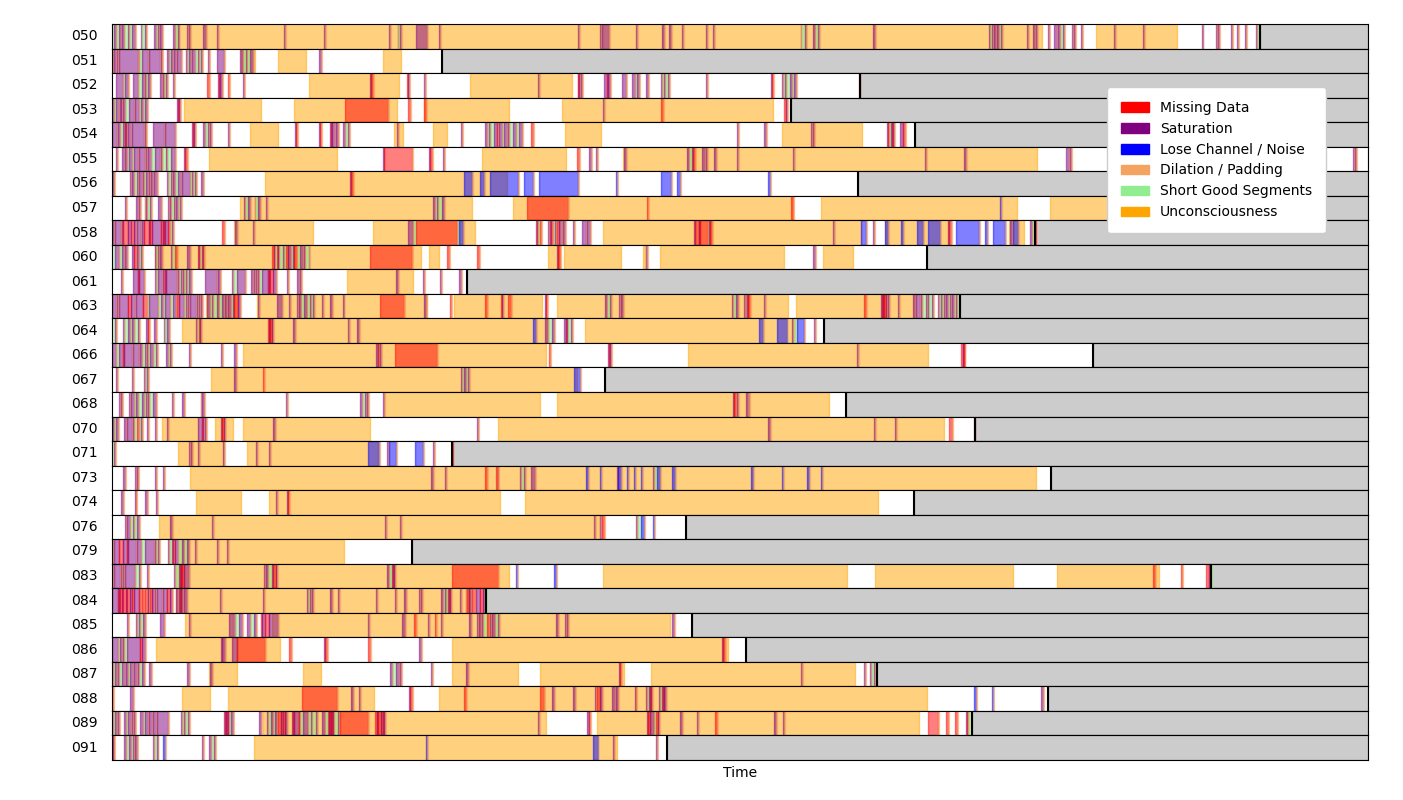}
    \caption{Overview of Individual Artifact Detection Results from a Subset of Records (1/3).}
    \label{fig:artifact_overview_1}
\end{figure}

\begin{figure}[h]
    \centering
    \includegraphics[width=\textwidth, trim=50 25 20 15, clip=TRUE]{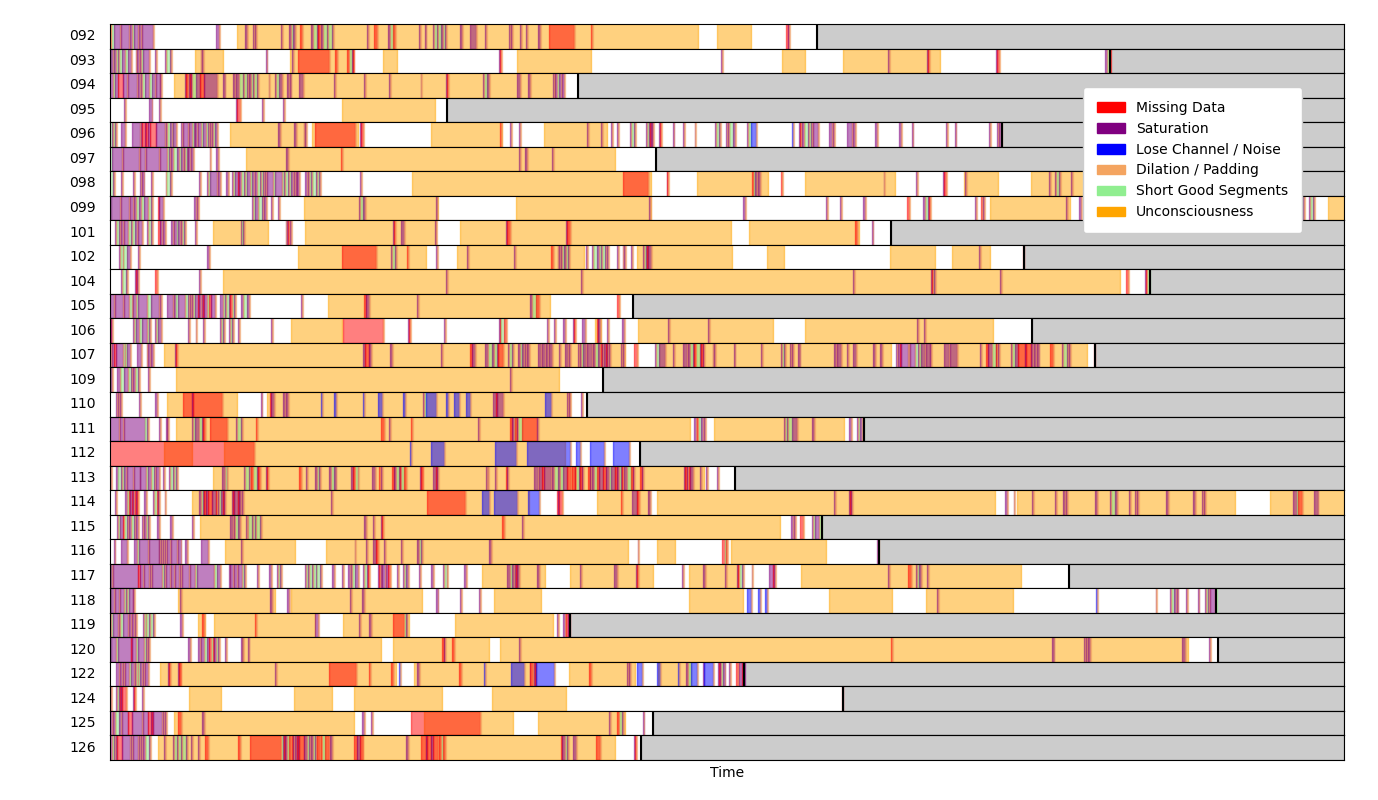}\\[-0.1cm]
    \includegraphics[width=\textwidth, trim=50 25 20 15, clip=TRUE]{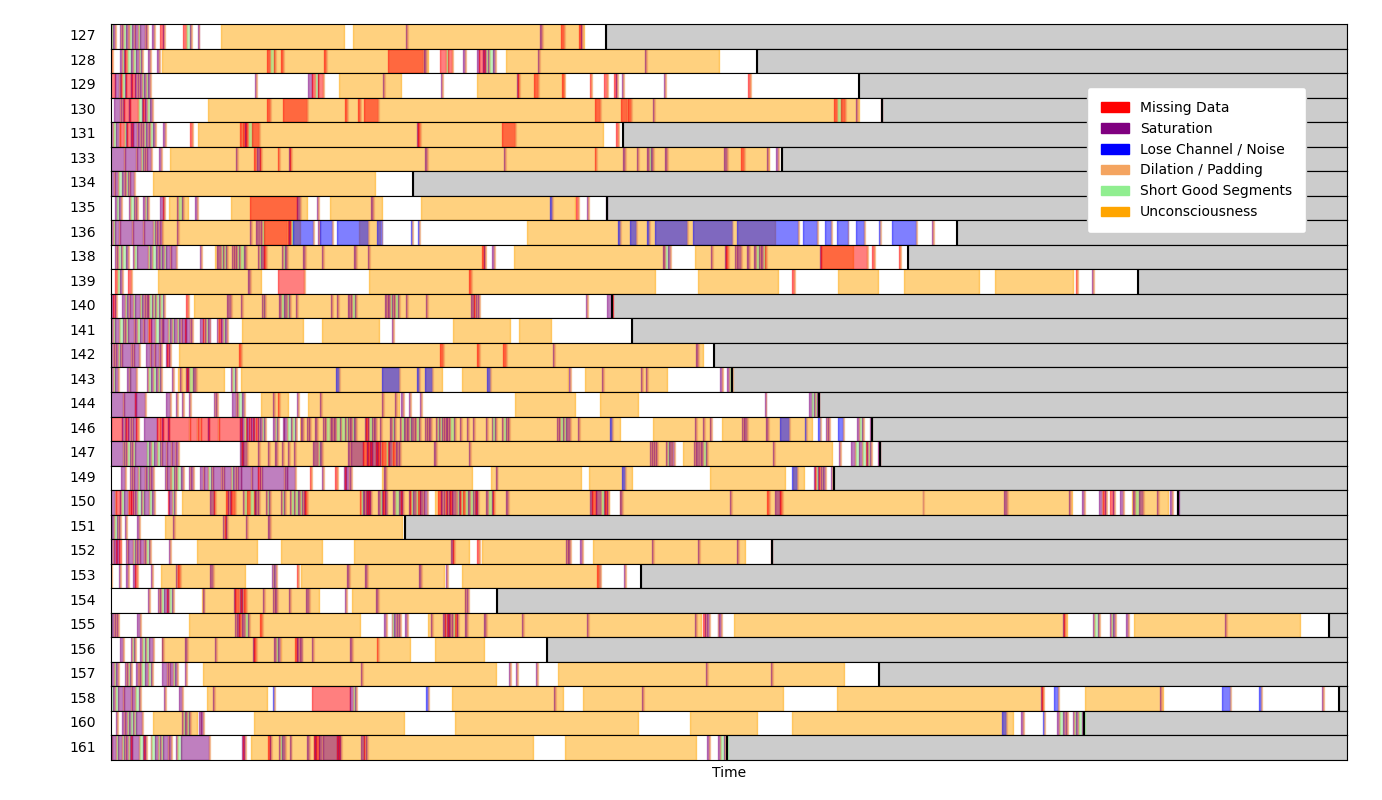}
    \caption{Overview of Individual Artifact Detection Results from a Subset of Records (2/3).}
    \label{fig:artifact_overview_2}
\end{figure}

\begin{figure}[h]
    \centering
    \includegraphics[width=\textwidth, trim=50 25 20 15, clip=TRUE]{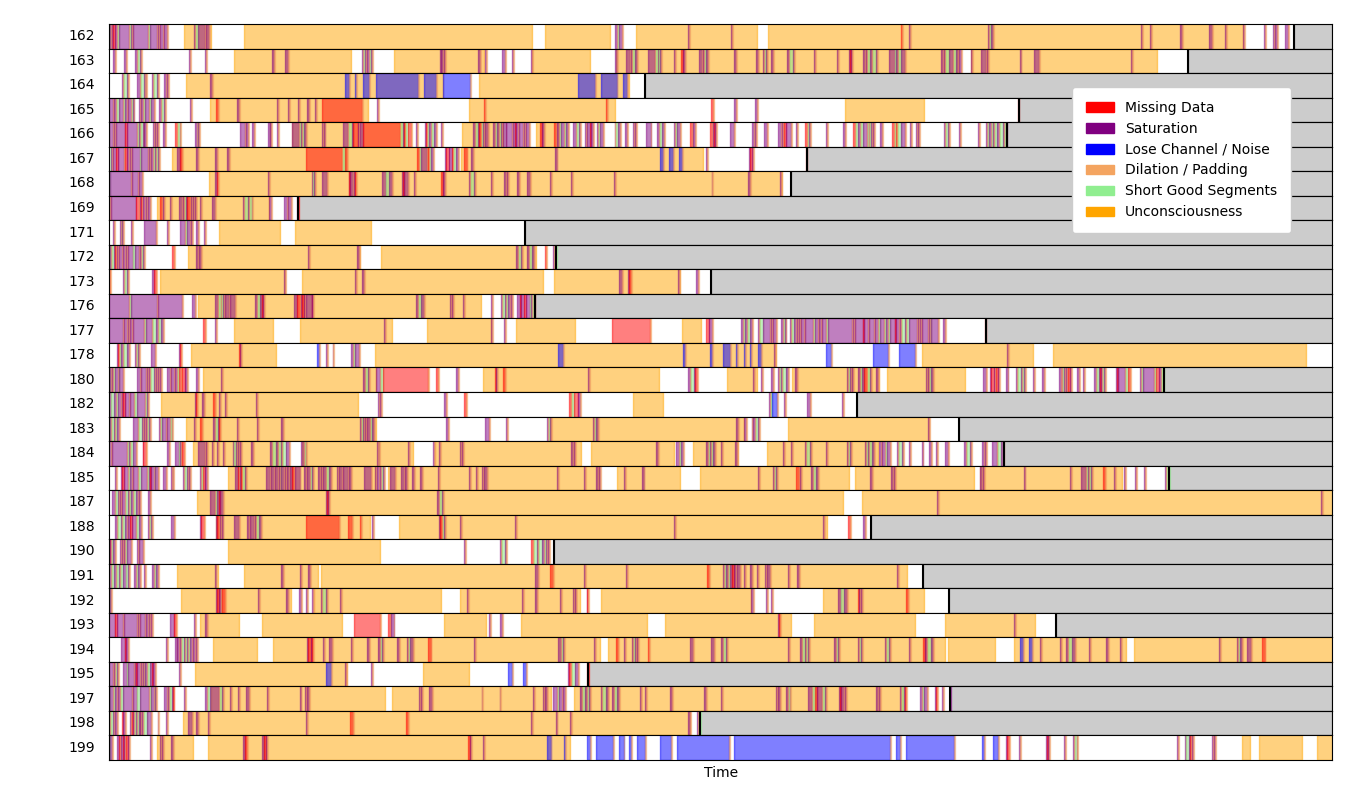}\\[-0.1cm]
    \includegraphics[width=\textwidth, trim=50 25 20 15, clip=TRUE]{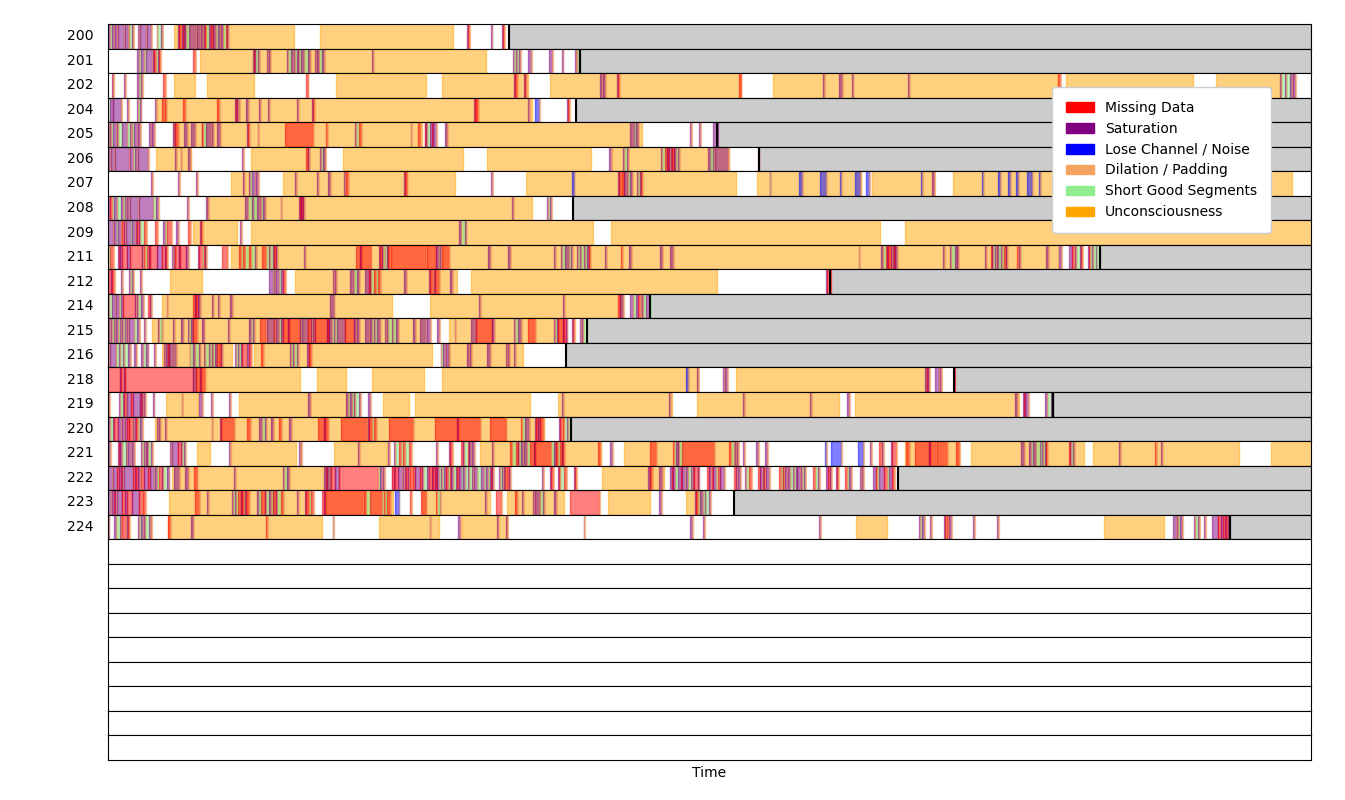}
    \caption{Overview of Individual Artifact Detection Results from a Subset of Records (3/3).}
    \label{fig:artifact_overview_3}
\end{figure}

Figure \ref{fig:artifact_overview_1} to \ref{fig:artifact_overview_3} give an overview of EEG signal quality on the record level. Researchers using this dataset should be aware of the unique environment the data was collected in.

\pagebreak



\clearpage

\section{Data Use Agreement}
\label{sec:data-use-agreement}

This Agreement governs the terms and conditions under which the Dataset (hereinafter "the Data") is provided to and utilized by the Data User (hereinafter "the User").

\begin{enumerate}
    \item \textbf{Anonymization and Current State of Technology.}
    The Data provided under this agreement has been subjected to anonymization procedures. All direct and indirect identifiers have been removed or transformed such that re-identification of individual data subjects is considered impossible under existing methodologies.

    \item \textbf{Risk Acknowledgment and Future Technical Evolution.} 
    While the Data is anonymized to the current technical standards, the User acknowledges that: The evolution of computational power and data-linking techniques may change the risk profile of the Data over time. It cannot be definitively guaranteed that future technological advancements will not enable the re-identification of individuals within the Dataset.

    \item \textbf{Prohibition of Re-identification.}
    In consideration of the residual risk mentioned in Section 2, the User explicitly agrees to the following:
    \begin{itemize}
        \item Non-Identification: The User shall not attempt to identify or contact any individual data subject included in the Data.
        \item No Re-identification Attempts: The User is strictly prohibited from employing any technical means, data-matching algorithms, or external data sources to reverse the anonymization process.
        \item Reporting: If the User inadvertently identifies any individual, they must immediately notify the Data Provider and cease use of the affected data points.
    \end{itemize}
    \item \textbf{Legal and Ethical Compliance}
    The User commits to processing and utilizing the Data exclusively within the following framework:
    \begin{itemize}
        \item Statutory Compliance: Use must strictly adhere to all applicable national and international laws, including but not limited to data protection regulations (e.g., GDPR, CCPA) and intellectual property laws.
        \item Ethical Standards: The Data shall be used solely for the purposes described in the research proposal or project description. Any use that contradicts established scientific ethical guidelines or harms the reputation of the data subjects is prohibited.
    \end{itemize}

    \item \textbf{Termination and Misuse.}
    Any violation of the provisions regarding re-identification or ethical compliance will result in the immediate termination of this license. The User may be held liable for legal consequences arising from the unauthorized processing of personal data resulting from a breach of this agreement.

\end{enumerate}

\pagebreak

\addcontentsline{toc}{section}{References}
\nocite{*}
\bibliographystyle{plainnat}
\bibliography{references}

\end{document}